\documentclass[useAMS,usenatbib]{mn2e}

\usepackage{amsmath,array,amsfonts}
\usepackage{amssymb}
\usepackage{graphicx}
\usepackage{natbib}
\usepackage{pstricks}
\usepackage{verbatim}
\usepackage{multirow}
\usepackage[shortlabels]{enumitem}
\usepackage{bbold}
\usepackage{pifont}
\usepackage{pstricks}
\usepackage{pst-pdf}
\usepackage{xcolor}

\newcommand{\ltsima} {$\; \buildrel < \over \sim \;$}
\newcommand{\gtsima} {$\; \buildrel > \over \sim \;$}
\newcommand{\lta} {\lower.5ex\hbox{\ltsima}}
\newcommand{\gta} {\lower.5ex\hbox{\gtsima}}

\def\sevem{{\tt SEVEM}}

\title[Radial derivatives as a test of pre-Big-Bang events]{Radial derivatives as a test of pre-Big-Bang events on the Planck data}

\author[R. Fern\'andez-Cobos et al.]{R. Fern\'andez-Cobos$^1$\thanks{e-mail:rfdz.cobos@gmail.com}, A. Marcos-Caballero$^{1,2}$, E. Mart\'inez-Gonz\'alez$^1$\\
$^1$     Instituto de F\'isica de Cantabria, CSIC-Universidad de Cantabria, Avda. de los Castros s/n, E-39005 Santander, Spain. \\
$^2$    Department of Theoretical Physics, University of the Basque Country, UPV/EHU, E-48080 Bilbao, Spain.}

\date{Accepted  Received ; in original form }

\begin{document}

\maketitle

\begin{abstract}
Although the search for azimuthal patterns in cosmological surveys is useful to characterise some effects depending exclusively on an angular distance within the standard model, they are considered as a key distinguishing feature of some exotic scenarios, such as bubble collisions or conformal cyclic cosmology (CCC). In particular, the CCC is a non-stardard framework which predicts circular patterns on the CMB intensity fluctuations. Motivated by some previous works which explore the presence of radial gradients, we apply a methodology based on the radial derivatives to the latest release of \textit{Planck} data. The new approach allows exhaustive studies to be performed at all sky directions at a HEALPix resolution of $N_{\mathrm{side}} = 1024$. Specifically, two different analyses are performed focusing on weight functions in both small (up to a $5$-degree radius) and large scales. We present a comparison between our results and those shown by \citet{An2018}, and \citet{An2018hwk}. In addition, a possible polarization counterpart of these circular patterns is also analysed for the most promising case. Taking into account the limitations to characterize the significance of the results, including the possibility of suffering a look-elsewhere effect, no strong evidence of the kind of circular patterns expected from CCC is found in the \textit{Planck} data for either the small or the large scales. 
\end{abstract}
\section{Introduction}
\label{sec:Introduction}
A usual way of testing the standard model of cosmology is to confront the data to alternative models and scenarios which provide a distinguishing prediction. In particular, looking for azimuthal patterns have been a recurrent topic in the literature, at least, since a spatial mapping of the cosmic microwave background (CMB) fluctuations is available. This symmetry is particularly convenient when modeling phenomena which depend exclusively on an angular distance, such as flows that extend homogeneously from a source, or spatial 2-point correlations. For this reason, azimuthal patterns are present in many different contexts, including some aspects within the standard cosmological paradigm, such as the monopolar contribution from the stacking of CMB peaks \citep[e.g.,][]{MarcosCaballero2016}, the imprint of Galactic supernova remnants \citep[cf.][]{Liu2014}, or the integrated Sachs-Wolfe effect from cosmic voids \citep[e.g.,][]{Martinez1990, Finelli2016}. Within more exotic scenarios, some specific azimuthal patterns have been considered, for instance, as a footprint of bubble collisions \citep[e.g.,][]{Aguirre2011}, the effect of a cosmic texture \citep{Cruz2007}, or an evidence of the intersection of different images of the same last-scattering surface in a closed topology \citep{Cornish1998}. 

Another scenario in which such circular patterns would be expected is the conformal cyclic cosmology (CCC) presented by \citet{Penrose2010}. Within this model, coalescences of black holes from the previous aeon would leave a particular mark on the CMB intensity fluctuations as low-variance azimuthal patterns. These imprints have been sought by different authors on the 7-year WMAP data. Firstly, \citet{Gurzadyan2010} claimed a $6\sigma$ detection of sky directions with anomalously low-variance angular profiles, although several subsequent papers denied that statistical significance. In particular, \citet{Wehus2011}, \citet{Moss2011}, and \citet{Hajian2011} showed that there is nothing special at the pointed centres of circles with anomalously low variance when considering the expected fluctuations in the whole data map. In a posterior paper, \citet{Gurzadyan2013} focused on the non-isotropic distribution of these low-variance concentric circles in the WMAP data. Such anisotropic distribution could be a reflect of the same effect that produces the already known CMB hemispherical asymmetry \citep[see][and references therein]{PlanckVII2018}, which is not necessarily related to any pre-Big-Bang phenomenology. In addition, \citet{DeAbreu2015} presented an analysis of both WMAP and \textit{Planck} data in which they demonstrated that the presence of these concentric circles are not significant with respect to the expected pattern from standard simulations. 

More recently, several papers have been produced in order to explore these circular patterns focusing on the analysis of  cumulative distribution functions (CDF) of some estimators computed from the data maps. For instance, \citet{Meissner2013} designed a filter which highlights the ring-type features up to scales associated with a radius of $22^{\circ}$ on the intensity CMB map. The significance of the most prominent structures are then evaluated in terms of the tails of the CDF of the filtered field. They claim a detection of ring-type structures on WMAP data ath the $99.7$ per cent confidence level. In addition, there are other two works focusing on \textit{Planck} data. In the first one, \citet{An2018} look for the presence of circles in the large-scale CMB field using the differences of the mean intensity between adjacent rings. They find an anomalous scale around $8^{\circ}$ of radius at a $99.6$ per cent confidence level, although they admit that this statistical significance is possibly overrated in light of a potential look-elsewhere effect. In a second paper, \citet{An2018hwk} explore the scales up to a $5$-degree radius computing the radial slope inside the rings, finding evidence of an anomalous detection at a $99.98$ per cent confidence level. Finally, in another recent paper, \citet{Jow2020} revisit the evidence of radial gradients, also including a polarization analysis. They conclude that the significance of these features is not relevant when the probability of finding a similar result within the whole range of scales is considered. 

In the present paper, we use a similar approach to characterize the circular patterns in terms of the radial derivatives. The method has already been used to analyse the CMB derivatives up to second order within discs by \citet{MarcosCaballero2017}. Since this methodology is more optimal than analogous approaches in real space, it allows exhaustive analyses to be performed at all sky directions at a HEALPix resolution of $\mathrm{N_{side}} = 1024$ \citep{Gorski2005}. Both the small and large scales of the latest \textit{Planck} data release are analysed in order to check the statistical significance of the results shown in previous works. In addition, a possible counterpart in polarization is also explored for the most promising case. In particular, a stacking analysis is performed to increase the signal-to-noise ratio of the polarization signal. The paper is structured as follows. The specific methodology is described in the next section. In Section~\ref{sec:Analysis}, we present a multi-scale analysis of azimuthal patterns in the \textit{Planck} data using the end-to-end \textit{Planck} simulations. The statistical significance of the results is discussed in Section~\ref{sec:Elsewhere}. Finally, Section~\ref{sec:Conclusions} collects the conclusions of the analysis.

%
 
\section{Methodology}
\label{sec:methodology}
In general, a vector field $\zeta^{\pm 1}$ on the sphere can be spanned in the helicity basis using the spin-weighted spherical harmonics ${}_{\pm 1}Y_{\ell m}$ as:
\begin{equation}
\label{eq:derivative}
\zeta^{\pm 1} \left(\mathrm{\mathbf{n}}\right) = \sum_{\ell = 1}^{\infty} \sum_{m=-\ell}^{\ell} \zeta_{\ell m}^{\pm 1}\ {}_{\pm 1}Y_{\ell m}\left(\mathrm{\mathbf{n}}\right),
\end{equation}
where $\zeta_{\ell m}^{\pm 1}$ denotes the coefficients of the expansion. Note that, in the particular case in which $\zeta^{\pm 1}$ is the gradient of a scalar field, these coefficients can be expressed in terms of the spherical harmonic coefficients $a_{\ell m}$ of the field as
\begin{equation}
\zeta_{\ell m}^{\pm 1} = \mp \sqrt{\dfrac{\left( \ell + 1 \right)!}{\left( \ell - 1 \right)!}} a_{\ell m}.
\end{equation}
  
\begin{figure}
  \center
\psset{unit=0.7cm}
\begin{pspicture}(2.5,2.5)(7.5,7.5)
\psdot*(5,5)
\psline[linewidth=1.2pt,linestyle=dotted]{->}(5,3)(6,3)
\psline[linewidth=1.2pt]{->}(5,3)(5,4)
\psline[linewidth=1.2pt,linestyle=dotted]{->}(5,7)(4,7)
\psline[linewidth=1.2pt]{->}(5,7)(5,6)
\psline[linewidth=1.2pt,linestyle=dotted]{->}(7,5)(7,6)
\psline[linewidth=1.2pt]{->}(7,5)(6,5)
\psline[linewidth=1.2pt,linestyle=dotted]{->}(3,5)(3,4)
\psline[linewidth=1.2pt]{->}(3,5)(4,5)
\psline[linewidth=1.2pt,linestyle=dotted]{->}(6.414213562373095, 6.414213562373095)(5.707106781186548, 7.121320343559643)
\psline[linewidth=1.2pt]{->}(6.414213562373095, 6.414213562373095)(5.707106781186548, 5.707106781186548)
\psline[linewidth=1.2pt,linestyle=dotted]{->}(3.585786437626905, 6.414213562373095)(2.878679656440357, 5.707106781186548)
\psline[linewidth=1.2pt]{->}(3.585786437626905, 6.414213562373095)(4.292893218813452, 5.707106781186548)
\psline[linewidth=1.2pt,linestyle=dotted]{->}(3.585786437626905, 3.585786437626905)(4.292893218813452, 2.878679656440357)
\psline[linewidth=1.2pt]{->}(3.585786437626905, 3.585786437626905)(4.292893218813452, 4.292893218813452)
\psline[linewidth=1.2pt,linestyle=dotted]{->}(6.414213562373095,3.585786437626905)(7.121320343559643, 4.292893218813452)
\psline[linewidth=1.2pt]{->}(6.414213562373095,3.585786437626905)(5.707106781186547, 4.292893218813452)
\pscircle[linecolor=gray,linestyle=dotted](5,5){0.9}
\pscircle[linecolor=gray,linestyle=dotted](5,5){2.6}
\rput(5.33,4.67){\psframebox*[]{$\mathrm{\mathbf{n}}$}}

\end{pspicture}
\caption{Flat projection of a small patch of the sphere where the locally-projected vector components are shown at some directions $\mathrm{\mathbf{n^\prime}}$ within a ring (gray dots) around the central direction $\mathrm{\mathbf{n}}$. The value of the weight function $W(\mathrm{\mathbf{n}}\cdot \mathrm{\mathbf{n^\prime}})$ is $1$ inside the ring, while it is set to zero outside this region. In the particular case in which the vector is the derivative of a scalar field, the integral of the tangential component (dotted arrows) inside the ring is null.}
\label{fig:spinor_components}
\end{figure}
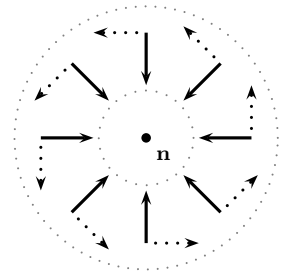

Suppose that the vector field $\zeta^{\pm 1}$ is expressed in Galactic coordinates. As we are interested in radial patterns from each sky direction $\mathrm{\mathbf{n}}$, we use a locally-defined rotation to align the local reference system of $\zeta^{\pm 1}$ at any other direction $\mathrm{\mathbf{n^\prime}}$ with the geodesic connecting $\mathrm{\mathbf{n}}$ and $\mathrm{\mathbf{n^\prime}}$. The field projected along the new axes (see Fig.~\ref{fig:spinor_components}) is then averaged using the corresponding weight function $W$ centred at each direction $\mathrm{\mathbf{n}}$. This weight function defines the region around $\mathrm{\mathbf{n}}$ in which each vector component is averaged, in such a way that it is null except inside the region, where its value is set to $1$. The following paragraphs are basically based on the methodology described in \citet{MarcosCaballero2017}. Explicitly, an estimator $\bar{\zeta}^{\pm 1}$ of the averaged vector components in the new locally-defined coordinates can be written in real space as:
\begin{equation}
\label{eq:mean_vector}
\bar{\zeta}^{\pm 1} \left( \mathrm{\mathbf{n}}\right) = \int {d^2 \mathrm{\mathbf{n^\prime}}\ W\left( \mathrm{\mathbf{n}}\cdot \mathrm{\mathbf{n^\prime}}\right) \zeta^{\pm 1}\left( \mathrm{\mathbf{n^\prime}}\right) e^{i\phi \left( \mathrm{\mathbf{n}}, \mathrm{\mathbf{n^\prime}}\right)}},
\end{equation}
where $\phi \left( \mathrm{\mathbf{n}}, \mathrm{\mathbf{n^\prime}}\right)$ is the angle between the geodesic connecting Galactic north and the $\mathrm{\mathbf{n^\prime}}$ direction, and the geodesic connecting the $\mathrm{\mathbf{n}}$ and $\mathrm{\mathbf{n^\prime}}$ directions. Note that, as the weight function presents azimuthal symmetry on the sphere with respect to the central direction, it depends only on the angular distance between the corresponding direction and the central point. 

Thereby, Eq.~\ref{eq:mean_vector} describes a complex scalar field. On the one hand, only the spinor components with even parity contribute to the real part of this field, defining the radial contribution. On the other hand, the odd-parity contributions give place to the imaginary part, which is the vector component in the direction orthogonal to the geodesic connecting $\mathrm{\mathbf{n}}$ and $\mathrm{\mathbf{n^\prime}}$. It is depicted using dotted arrows in Fig.~\ref{fig:spinor_components}. Insofar as we are interested in the CMB derivatives, the integral of this tangential component within the azimuthally-symetric region selected by $W$ is necessarily null. This is because, in this case (as well as for all the greater order derivatives), it represents the curl contribution of a gradient. Therefore, an estimator $\bar{\eta} \equiv \bar{\zeta}^{+1} = \bar{\zeta}^{-1}$ for the averaged radial derivative of the CMB within the region defined by $W$ around each direction $\mathrm{\mathbf{n}}$ can be expanded in terms of the standard spherical harmonics $Y_{\ell m}$ as follows: 

\begin{equation}
\label{eq:radial_derivative}
\bar{\eta} \left( \mathrm{\mathbf{n}}\right) = \sum_{\ell= 1}^\infty {\sum_{m=-\ell}^{\ell}{D_{\ell}a_{\ell m}Y_{\ell m}\left( \mathrm{\mathbf{n}}\right)}},
\end{equation}
where $a_{\ell m}$ are the spherical harmonic coefficients of the intensity field. Eq.~\ref{eq:radial_derivative} implies that both the radial projection and the average within the azimuthally-invariant region defined by $W$ can be encoded as a convolution of the CMB intensity map with a window function $D_{\ell}$. This filter is computed as follows:
\begin{equation}
D_{\ell} = \sum_{\ell^\prime = 0}^\infty {M_{\ell \ell^\prime} W_{\ell^\prime}},
\end{equation} 
where $W_\ell$ denotes the filter coefficients of the weight function $W$, and the coupling matrix can be written as:
\begin{eqnarray}
M_{\ell \ell^\prime} & = & -\left( 2\ell^\prime + 1\right) \sqrt{\dfrac{\left(\ell + 1 \right)!}{\left(\ell - 1 \right)!}} \sum_{L = |\ell - \ell^\prime|}^{\ell + \ell^\prime}\left( 2L + 1\right) \sqrt{\dfrac{\left(L - 1 \right)!}{\left(L + 1 \right)!}} \times \nonumber \\
 & & \times \left( {\begin{array}{ccc} \ell & \ell^\prime & L \\ 0 & 0 & 0 \end{array}}\right) \left( {\begin{array}{ccc} \ell & \ell^\prime & L \\ 1 & 0 & -1 \end{array}}\right)c_L,
\end{eqnarray}
with
\begin{equation}
c_{L} = \left\lbrace \begin{array}{cc} \dfrac{\pi}{2^{\left(L+3\right)}} \dfrac{\left( L + 1\right)}{L}\left({\begin{array}{c} L + 1 \\ \\ \dfrac{L+1}{2}\end{array}} \right)^2 &,\ \mathrm{if}\ L\ \mathrm{odd} \\  \\ 0 &,\ \mathrm{if}\ L\ \mathrm{even} \end{array}\right..
\end{equation}

In practice, these $c_L$ coefficients are computed using the following recursive rule:
\begin{equation}
c_{L+2} = \dfrac{L\left( L + 2\right)}{\left( L + 3\right)\left( L + 1\right)} c_L,
\end{equation}
with $c_0 = 0$, and $c_1 = \pi/4$.

Explicitly, in the case in which $W$ selects a disc with angular radius $\epsilon$ around the direction $\mathrm{\mathbf{n}}$, the harmonic coefficients $W_\ell$ can be computed as:
\begin{equation}
\label{eq:wl_disc}
W_{\ell}^{\mathrm{disc}}(\mu) = \left\lbrace \begin{array}{cc} 1 &,\ \mathrm{if}\ \ell = 0 \\ \\ 
-\sqrt{\dfrac{1+\mu}{1-\mu}} \dfrac{P_{\ell}^{1}(\mu)}{\ell (\ell + 1)} &, \ \mathrm{if}\ \ell \ne 0  \end{array}\right.,
\end{equation}
where $\mu = \cos{\epsilon}$, and $P_{\ell}^{1}(\mu)$ denotes the associated Legendre polynomial with $m = 1$. The coefficients for a ring with inner angular radius $r$ and thickness $\epsilon$ are computed as a normalized difference of two discs:
\begin{equation}
\label{eq:wl_ring}
W_{\ell}^{\mathrm{ring}}(\mu_{\mathrm{i}}, \mu_{\mathrm{o}}) = \dfrac{\left( 1 - \mu_{\mathrm{o}} \right)W_{\ell}^{\mathrm{disc}}(\mu_{\mathrm{o}}) -  \left( 1 - \mu_{\mathrm{i}} \right)W_{\ell}^{\mathrm{disc}}(\mu_{\mathrm{i}})}{\mu_{\mathrm{i}} - \mu_{\mathrm{o}}},
\end{equation}
with inner $\mu_{\mathrm{i}} = \cos{r}$, and outer $\mu_{\mathrm{o}} = \cos{\left(r+\epsilon\right)}$ scales.

\begin{figure}
\includegraphics[scale=0.55]{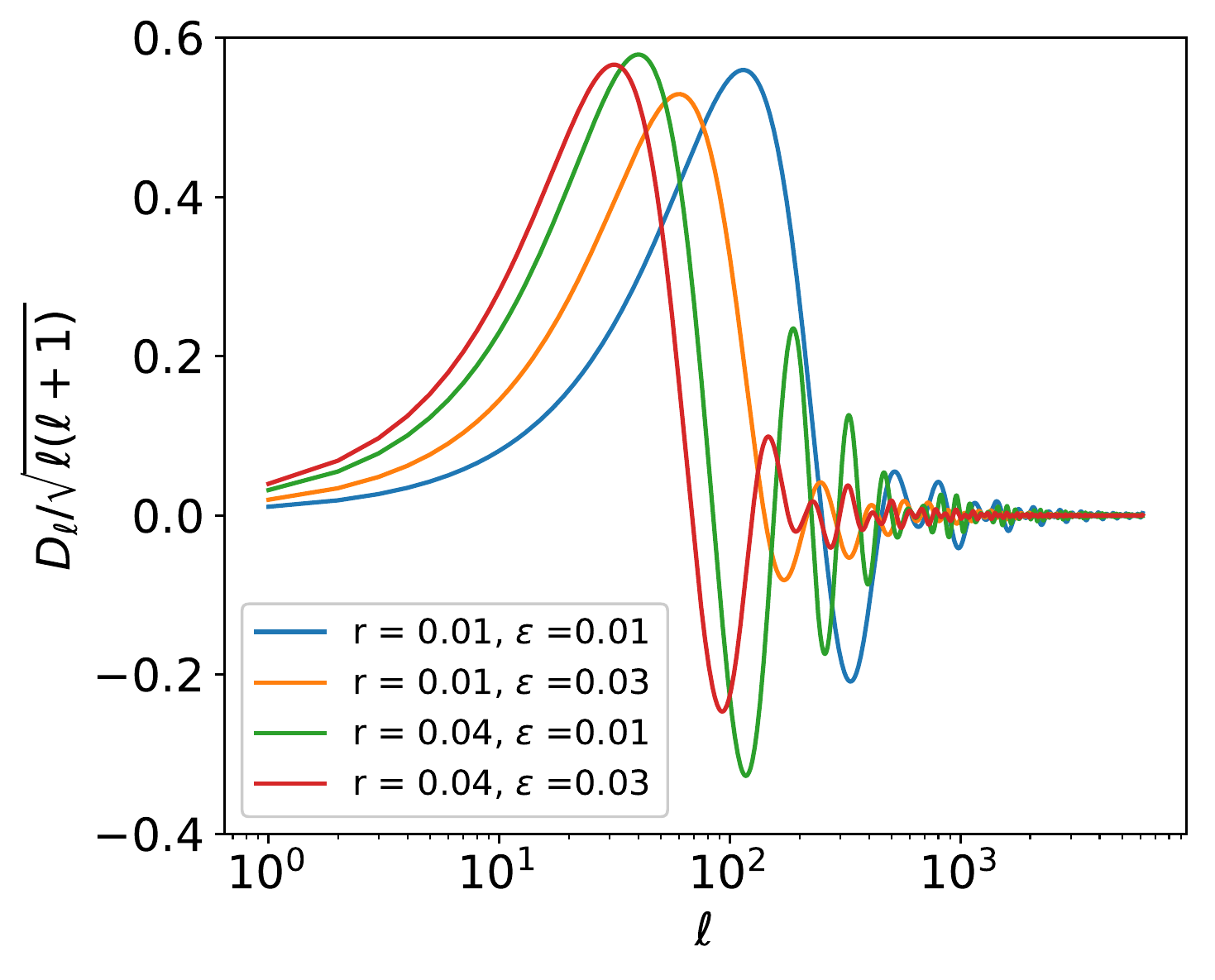} \\
\includegraphics[scale=0.55]{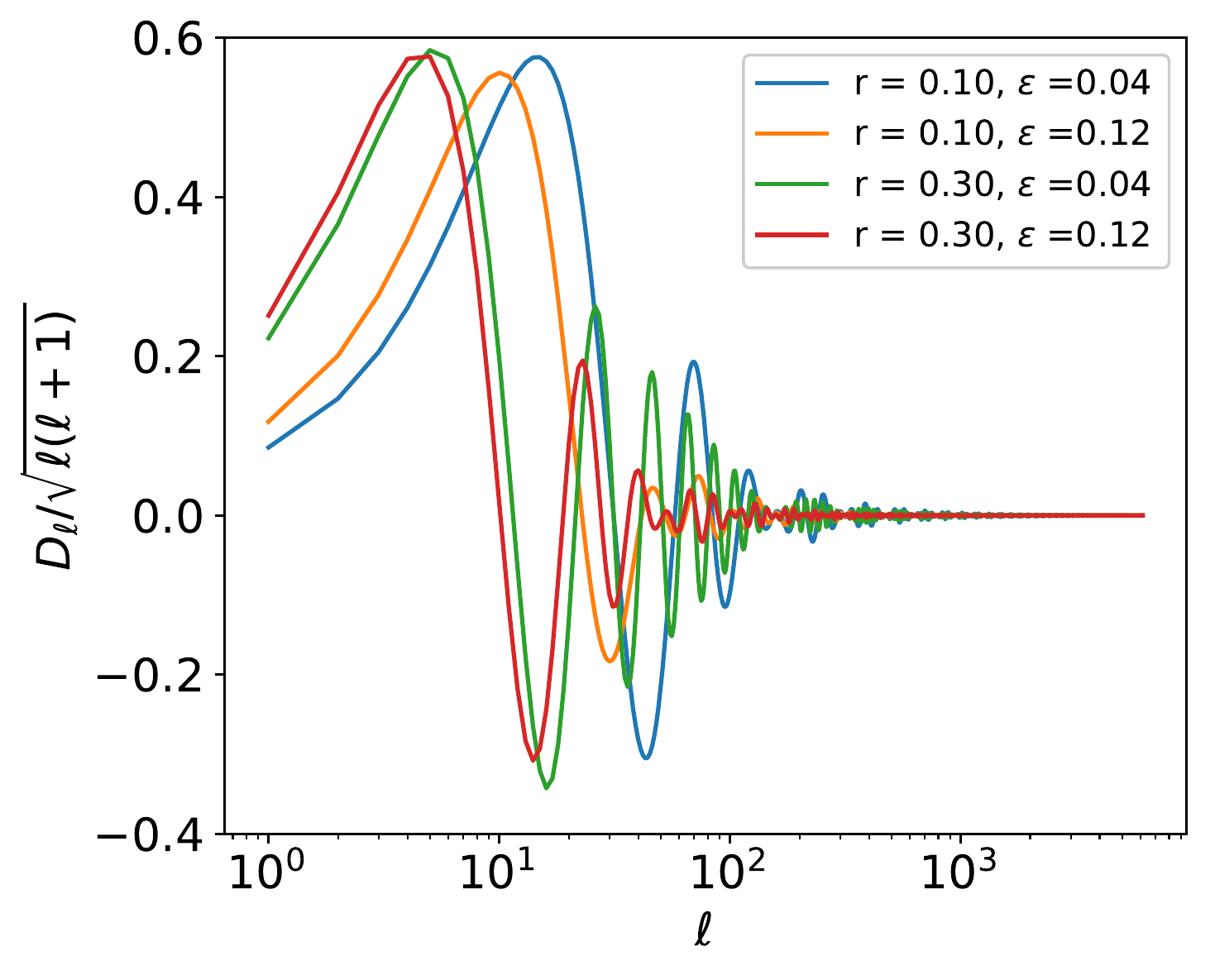}
\caption{Examples of $D_{\ell}$ filters for the small-scale (upper panel) and the large-scale (lower panel) regimes. The $W$ functions are parameterized by the inner radius $r$, and the thickness of the ring $\epsilon$. Both angular distances are expressed in radians.}
\label{fig:filters}
\end{figure}

Summarizing, within this framework, it is possible to obtain a map of averaged radial gradients $\bar{\eta}$ by computing a single convolution of the data map. Figure~\ref{fig:filters} shows some examples of $D_{\ell}$ filters for both the small-scale (upper panel) and the large-scale (bottom panel) regimes considered in the following sections. Therefore, this approach is less computationally expensive than the equivalent real-space methodologies. For instance, while a coarse-grained grid of centres is selected in other analyses, such as \citet{An2018}, this methodology enables us to consider the radial gradient around all pixels of the data map. 

%

\section[\textit{Planck} data analysis]{\textit{Planck} data analysis}
\label{sec:Analysis}

In this section, we analyse the foreground-cleaned CMB maps provided as part of the \textit{Planck} 2018 data release \citep{PlanckIV2018}. In order to compute the significance of the results, the same analysis is applied both to data and FFP10\footnote{The full focal-plane (FFP) simulations are Monte Carlo realizations which include the CMB signal, the instrumental noise, and the systematic effects expected in the \textit{Planck} data. See \citet{PlanckIV2018} and references therein for further details. All simulated maps used in this paper have been previously propagated through the corresponding component-separation pipeline by the \textit{Planck} Collaboration as described in \citet{PlanckVII2018}.} simulations. These end-to-end realizations are generated with the best available model for the anisotropic instrumental noise, beam asymmetries and systematic effects present in the the data. All maps are convolved with a Gaussian beam of $10^\prime$ FWHM at a HEALPix resolution of $\mathrm{N_{side}} = 1024$. Note that,  while $1000$ CMB realizations have been provided, only $300$ instrumental noise realizations are available. Following the methodology used in \citet{PlanckVII2018}, we permute the signal and noise realizations to generate a total of $999$ simulations. In particular, we add the same $300$ noise realizations to the $0$-$299$, $300$-$599$ and $600$-$899$ sets, and the first $100$ noise realizations to the $900$-$999$ CMB simulated maps\footnote{Except for the CMB realization $970$, because, as noted in \citet{PlanckVII2018}, the simulated map is corrupted.}. The whole analysis is performed using the \sevem\ maps, although we have checked that similar results are obtained from the other \textit{Planck} foreground-removed CMB maps.

Different sizes for the $W$ function are considered in terms of the inner radius $r$ and the thickness $\epsilon$. In the present work, two different analyses are distinguished. On the one hand, a small-scale analysis in which both quantities are sampled at intervals of $0.01$~rad, so that the thickness does not exceed $2.5$ degrees and the outer radius is always smaller than $5$ degrees. On the other hand, we explore a large-scale domain from $r=0.06$~rad to $0.34$~rad with intervals of $0.04$~rad. Three different ring widths are considered in this case, namely $\epsilon = 0.04$~rad, $0.08$~rad, and $0.12$~rad. 

Given a certain weight function, the following steps are performed for both data and simulated maps:
\begin{enumerate}[leftmargin=.75cm, label=\roman*)]
\item Monopole and dipole are removed outside the \textit{Planck} confidence mask.
\item The CMB intensity map is convolved with the corresponding $D_\ell$ according to Eq.~\ref{eq:radial_derivative}. 
\item The $\bar{\eta}$ map is then normalized by the pixel-dependent standard deviation computed from the first $900$ simulated maps.
\end{enumerate}
This normalization enables to weight properly those pixels mitigated by the null values from the mask. Although no mask apodization is applied before convoluting, this seems to be precise enough in the large-scale case, in which the \textit{Planck} confidence mask is kept in the $\bar{\eta}$ map. We avoid that potencial mask effects bias the results applying the same procedure to both data and simulations. However, in the analysis of the small scale, we can afford to adopt a more conservative approach. To ensure that the convolution does not take into account any unreliable region, the \textit{Planck} confidence mask is extended in the following way: we convolve the mask with a Gaussian filter in which the outer radius of $W$ is taken as FWHM, and then we set to zero all values less than $0.9$. The $\bar{\eta}$ map is then remasked with the extended mask. This approach prevents point sources from growing too much. Nevertheless, we check for the considered scales that similar results are obtained from a more conservative mask in which all centers whose $W$ function overlaps more than a $1$ per cent are excluded. 

The following estimators are then computed from the data allowed by the corresponding mask:
\begin{enumerate}[leftmargin=.75cm, label=\roman*)]
\item[iv-a)] The extrema (maxima and minima) above different thresholds from the normalized $\bar{\eta}$ map. 
\item[iv-b)] The CDF from the normalized $\bar{\eta}$ map is sampled by using $10000$ bins. We check that the results remain basically unchanged when a greater number of bins is considered.
\end{enumerate}

Finally, in order to assess the statistical significance of the results, we compute the corresponding $p$-values taking as reference the distribution of values obtained from the FFP10 simulations.

\subsection{Number of extrema}
\label{subsec:extrema}
The number of maxima $N_{\land}$ (minima $N_{\lor}$) above (below) different thresholds obtained from the data is  compared with the values computed from $900$ FFP10 simulations. This comparison is made in terms of a non-standard $p$-value defined as $P(N_{\land/\lor}>N_{\land/\lor}^{\mathrm{data}})$, namely the probability of finding a number of independent extrema strictly greater than that obtained from the data, given the standard model. In addition, we check that very few extrema are expected above $4.5\sigma$. The fact that it is common to find realizations in which there is no extreme greater than this threshold is the reason for not taking the standard definition of a $p$-value. If we considered $P(N_{\land/\lor}\geq N_{\land/\lor}^{\mathrm{data}})$, i.e. the probability of finding a number of extrema greater or equal than that obtained from the data, then the $p$-value would saturate provided that no extrema above this threshold are obtained from the data. As expected, the difference between both definitions is not significant at low thresholds ($3\sigma$ and $3.5\sigma$), while the former yields smaller $p$-values at high thresholds ($4\sigma$ and $4.5\sigma$).

For the small-scale analysis, it is expected some clustering of extrema in particular regions. Mainly for visualization purposes, we additionally select an independent sample keeping only the most extreme one from a region within the correlation scale given by twice the outer radius of the ring. On the contrary, such criterion makes little sense for the large-scale analysis since the proportion of overlapping area of the $W$ functions is smaller. In addition, as the number of extrema above the considered thresholds is smaller for this case, a reduction of the sample implies poorer statistics.
 
\subsubsection*{Small-scale analysis}
At a HEALPix resolution of $\mathrm{N_{side}} = 1024$, we find no significant differences at small scales between considering the whole sample of extrema or selecting the independent ones. As it is preferable to consider the independent sample to identify different regions in the sky, the figures in the present section are based on this subset of extrema for a better visualization. Accordingly, avoiding being redundant, only $p$-values for independent extrema are shown to allow a consistent comparison with the figures. We show the results for these $p$-values in Table~\ref{tab:pvalues_extrema_gt_ind} for different small-scale weight functions with inner radius $r$ and thickness $\epsilon$. In general, the number of extrema seems to be compatible with the expected value from the \textit{Planck} model. An exception is found for a weight function with $r = 0.01$~rad and $\epsilon = 0.02$~rad. In this case, the number of maxima above $4.5\sigma$ obtained from the data is, in fact, strictly greater than the same number computed from any of the FFP10 simulations. However, using the standard definition of the $p$-value, the statistical significance drops when we focus on maxima above $4\sigma$, in which case we obtain $5$ times the probability shown in the table.

\begin{table*}
\begin{center}
\resizebox{\textwidth}{!} {
\begin{tabular}{|cc|cccc|cccc|}
\hline
\multicolumn{2}{|c|}{Weight functions} & \multicolumn{4}{c|}{$P$-values for maxima} & \multicolumn{4}{c|}{$P$-values for minima} \\
$r$ (rad) & $\epsilon$ (rad)  & $>3.0\sigma$ & $>3.5\sigma$ & $>4.0\sigma$ & $>4.5\sigma$ & $<-3.0\sigma$ & $<-3.5\sigma$ & $<-4.0\sigma$ & $<-4.5\sigma$\\
\hline
\hline
\multirow{4}{*}{0.00} & 0.01 & $0.751$ & $0.371$ & $0.176$ & $0.338$ & $0.019$ & $0.082$ & $0.349$ & $0.742$ \\
					  & 0.02 & $0.066$ & $0.130$ & $0.083$ & $0.828$ & $0.439$ & $0.800$ & $0.953$ & $0.843$ \\ 
					  & 0.03 & $0.350$ & $0.120$ & $0.034$ & $0.037$ & $0.462$ & $0.036$ & $0.334$ & $0.108$ \\
					  & 0.04 & $0.567$ & $0.510$ & $0.724$ & $0.253$ & $0.039$ & $0.074$ & $0.051$ & $0.598$ \\	
\hline
\multirow{4}{*}{0.01} & 0.01 & $0.130$ & $0.051$ & $0.008$ & $0.074$ & $0.272$ & $0.349$ & $0.818$ & $0.696$ \\
					  & 0.02 & $0.448$ & $0.150$ & $0.001$ & $<0.001$ & $0.280$ & $0.040$ & $0.271$ & $0.360$ \\ 
					  & 0.03 & $0.603$ & $0.329$ & $0.174$ & $0.272$ & $0.324$ & $0.002$ & $0.180$ & $0.624$ \\
					  & 0.04 & $0.827$ & $0.958$ & $0.238$ & $0.560$ & $0.069$ & $0.113$ & $0.033$ & $0.190$ \\	
\hline
\multirow{4}{*}{0.02} & 0.01 & $0.350$ & $0.232$ & $0.032$ & $0.007$ & $0.011$ & $0.073$ & $0.021$ & $0.094$ \\
					  & 0.02 & $0.127$ & $0.050$ & $0.157$ & $0.056$ & $0.032$ & $0.264$ & $0.307$ & $0.754$ \\ 
					  & 0.03 & $0.637$ & $0.870$ & $0.890$ & $0.272$ & $0.061$ & $0.350$ & $0.309$ & $0.262$ \\
					  & 0.04 & $0.387$ & $0.778$ & $0.983$ & $0.564$ & $0.194$ & $0.544$ & $0.148$ & $0.183$ \\	
\hline
\multirow{4}{*}{0.03} & 0.01 & $0.090$ & $0.016$ & $0.019$ & $0.426$ & $0.588$ & $0.761$ & $0.341$ & $0.090$ \\
					  & 0.02 & $0.321$ & $0.293$ & $0.441$ & $0.368$ & $0.242$ & $0.363$ & $0.283$ & $0.376$ \\ 
					  & 0.03 & $0.166$ & $0.811$ & $0.616$ & $0.266$ & $0.286$ & $0.548$ & $0.188$ & $0.080$ \\
					  & 0.04 & $0.077$ & $0.349$ & $0.373$ & $0.544$ & $0.189$ & $0.188$ & $0.400$ & $0.530$ \\	
\hline
\multirow{4}{*}{0.04} & 0.01 & $0.537$ & $0.659$ & $0.186$ & $0.080$ & $0.010$ & $0.018$ & $0.187$ & $0.227$ \\
					  & 0.02 & $0.120$ & $0.046$ & $0.048$ & $0.160$ & $0.141$ & $0.461$ & $0.287$ & $0.138$ \\ 
					  & 0.03 & $0.740$ & $0.619$ & $0.420$ & $0.248$ & $0.414$ & $0.438$ & $0.303$ & $0.250$ \\
					  & 0.04 & $0.136$ & $0.052$ & $0.564$ & $0.563$ & $0.550$ & $0.091$ & $0.223$ & $0.043$ \\	
\hline
\end{tabular}
}
\end{center}
\caption{Probability of finding a number of independent $\bar{\eta}$ extrema in FFP10 simulations which is strictly greater than that obtained from the \sevem\ data for different thresholds and small-scale weight functions with inner radius $r$ and thickness $\epsilon$.}
\label{tab:pvalues_extrema_gt_ind}
\end{table*}

\begin{figure*}
\centering{
\includegraphics[scale=0.7]{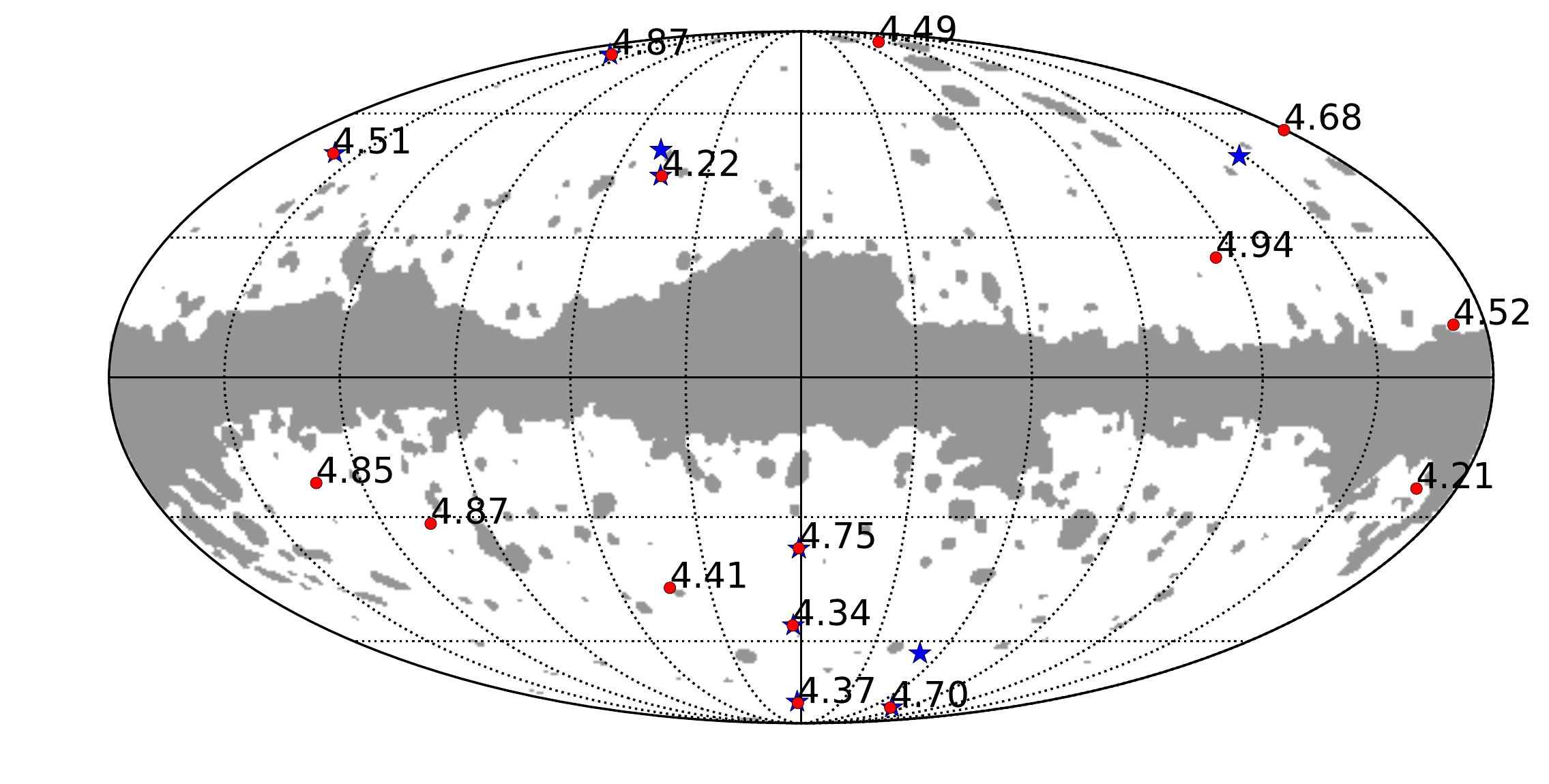} 
}
\caption{The $15$ most prominent maxima of the radial gradient averaged with a weight function with $r=0.01$~rad and $\epsilon=0.02$~rad (red points). The numbers in the map correspond with the value of the normalized $\bar{\eta}$ map at the corresponding direction. We also plot the $10$ directions given for this case in \citet{An2018hwk} as blue stars. As background, it is shown the extended mask for this case.}
\label{fig:directions}
\end{figure*}

This scale is also highlighted as possibly anomalous by \citet{An2018hwk}. Note that the sign convention is changed between that work and the present analysis. As shown in Figure~\ref{fig:spinor_components}, we consider that a radial gradient which points towards the centre is positive, while the convention used by these authors takes a slope as positive when it points outwards the centre. The red points in Figure~\ref{fig:directions} depict the sky directions in which the $15$ most prominent independent extrema of $\bar{\eta}$ are found for this case. The numbers in the map correspond with the value of the normalized $\bar{\eta}$ map at the corresponding central direction. We also plot the $10$ directions given for the same weight function by \citet{An2018hwk} as blue stars, $7$ of which are among our $15$ sky directions. However, there are only $4$ matches within our $10$ most prominent maxima. We have checked that considering a grid of centres at $\mathrm{N_{side}} = 64$ (as the mentioned authors do) does not make a big difference in our analysis. Other variations between both methodologies might explain these discrepancies. In the first place, although they are supposed to trace the same, we use different estimators in practice. They might involve different kinds of numerical errors. For instance, as can be seen in Figure~\ref{fig:filters}, the $D_{\ell}$ filters, specially for small scales, are not bandlimited. Nevertheless, this effect is expected to be innocuous to the conclusions of the analysis. More important is the use of the FFP10 end-to-end simulations as a guarantee that the noise properties and some systematics present in the \textit{Planck} data are taken into account in the simulated maps. In addition, the normalization of the radial derivative by the pixel-by-pixel standard deviation to ensure that all pixels are properly weighted may produce changes in the results.

\begin{figure*}
\centering{
\includegraphics[scale=0.3]{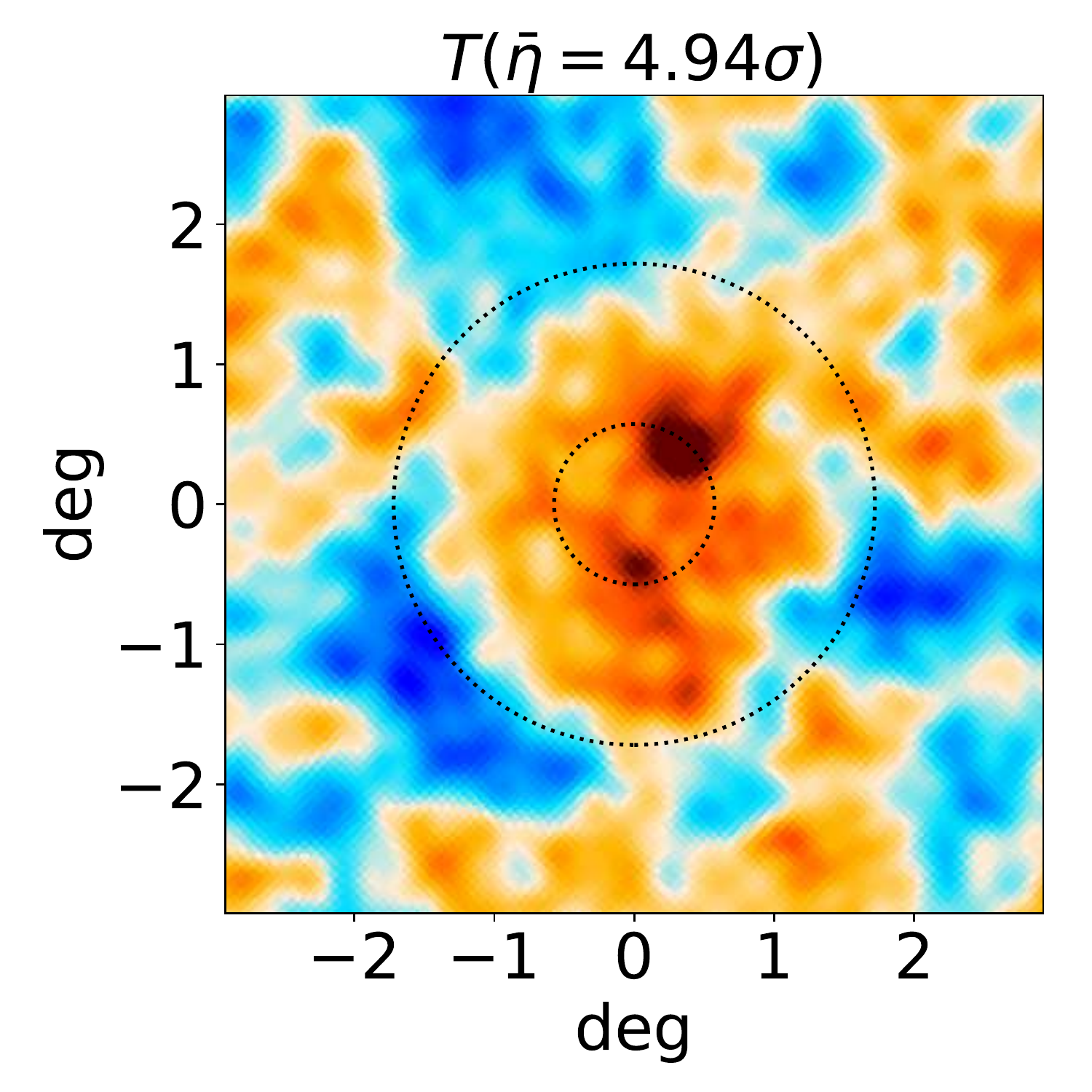} 
\includegraphics[scale=0.3]{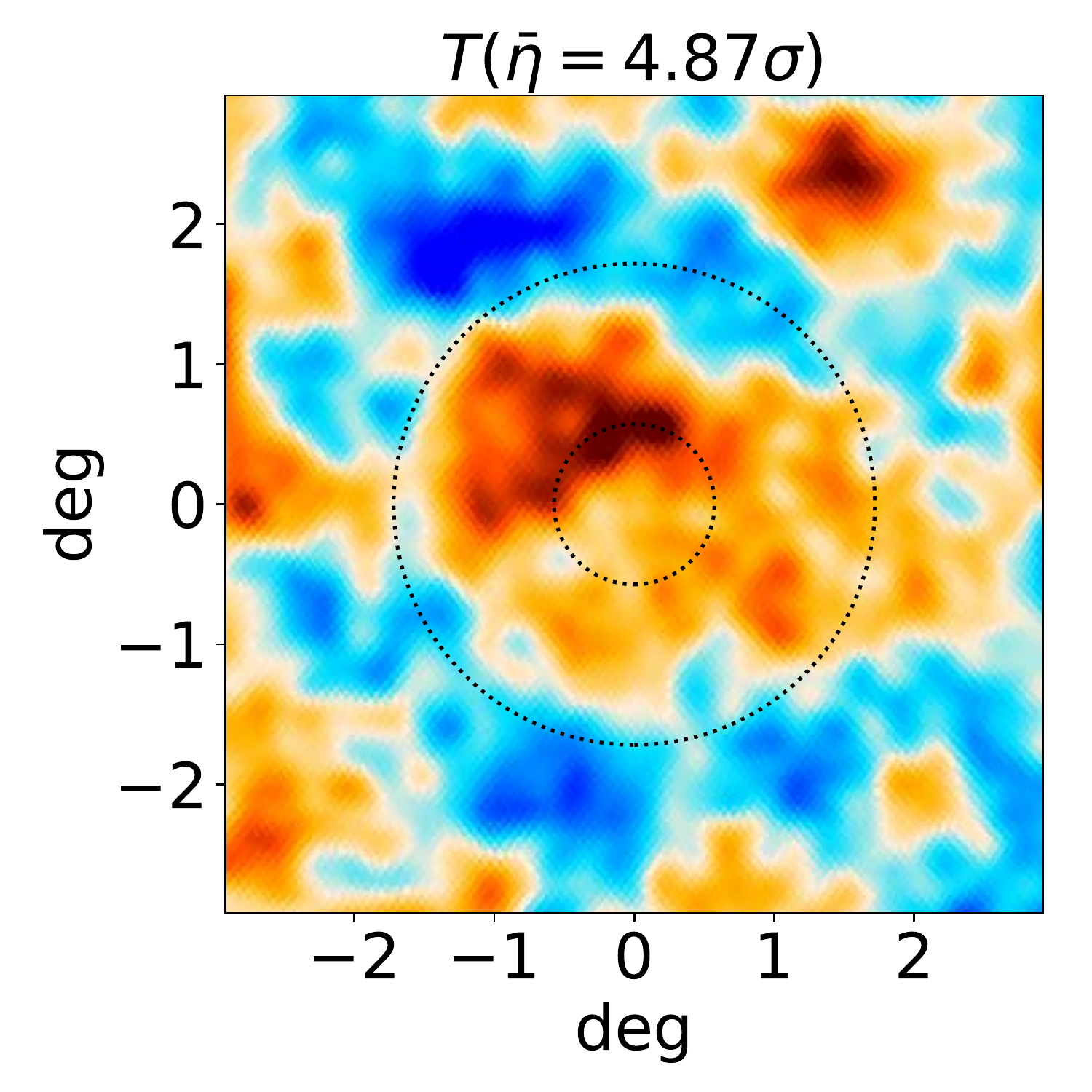} 
\includegraphics[scale=0.3]{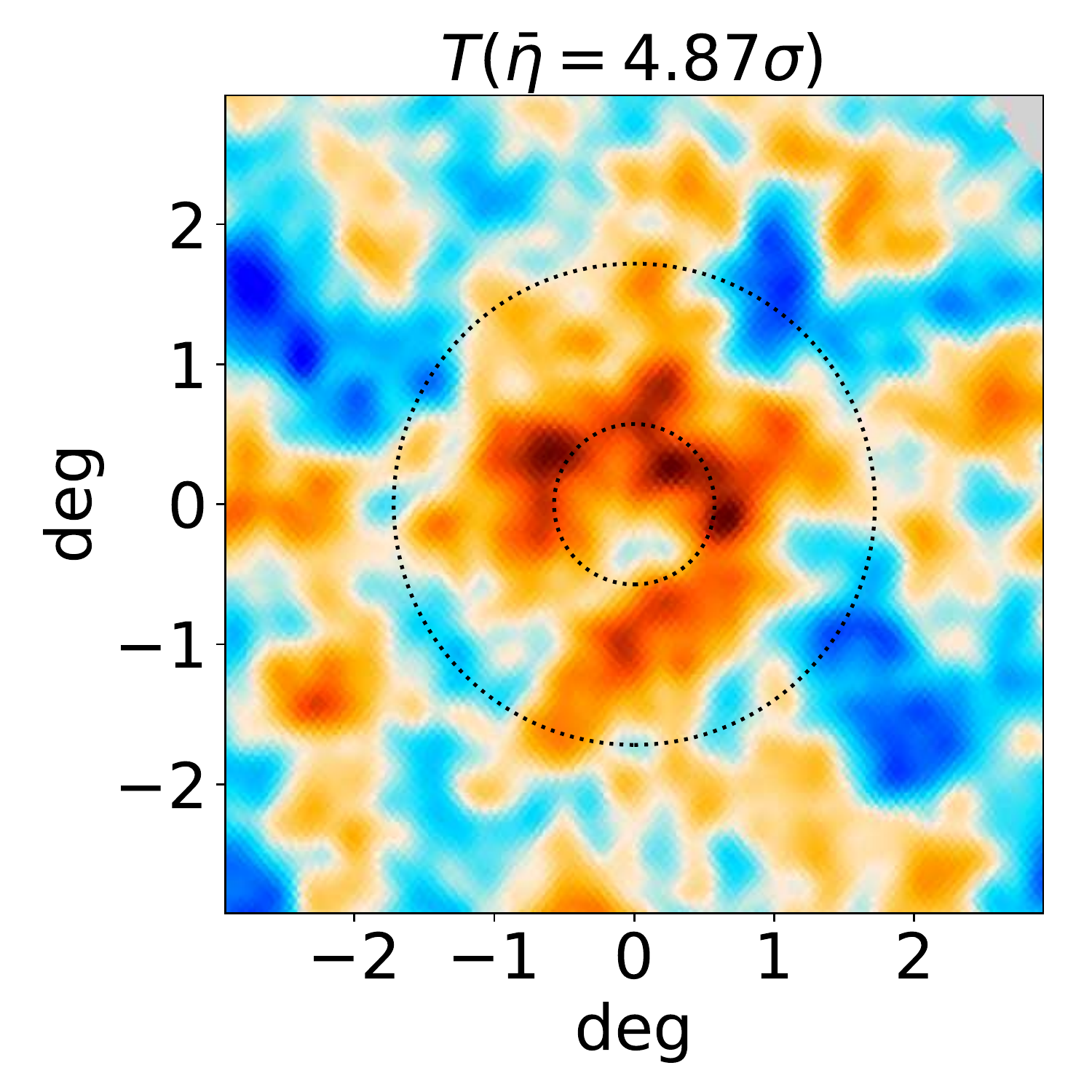} \\
\includegraphics[scale=0.3]{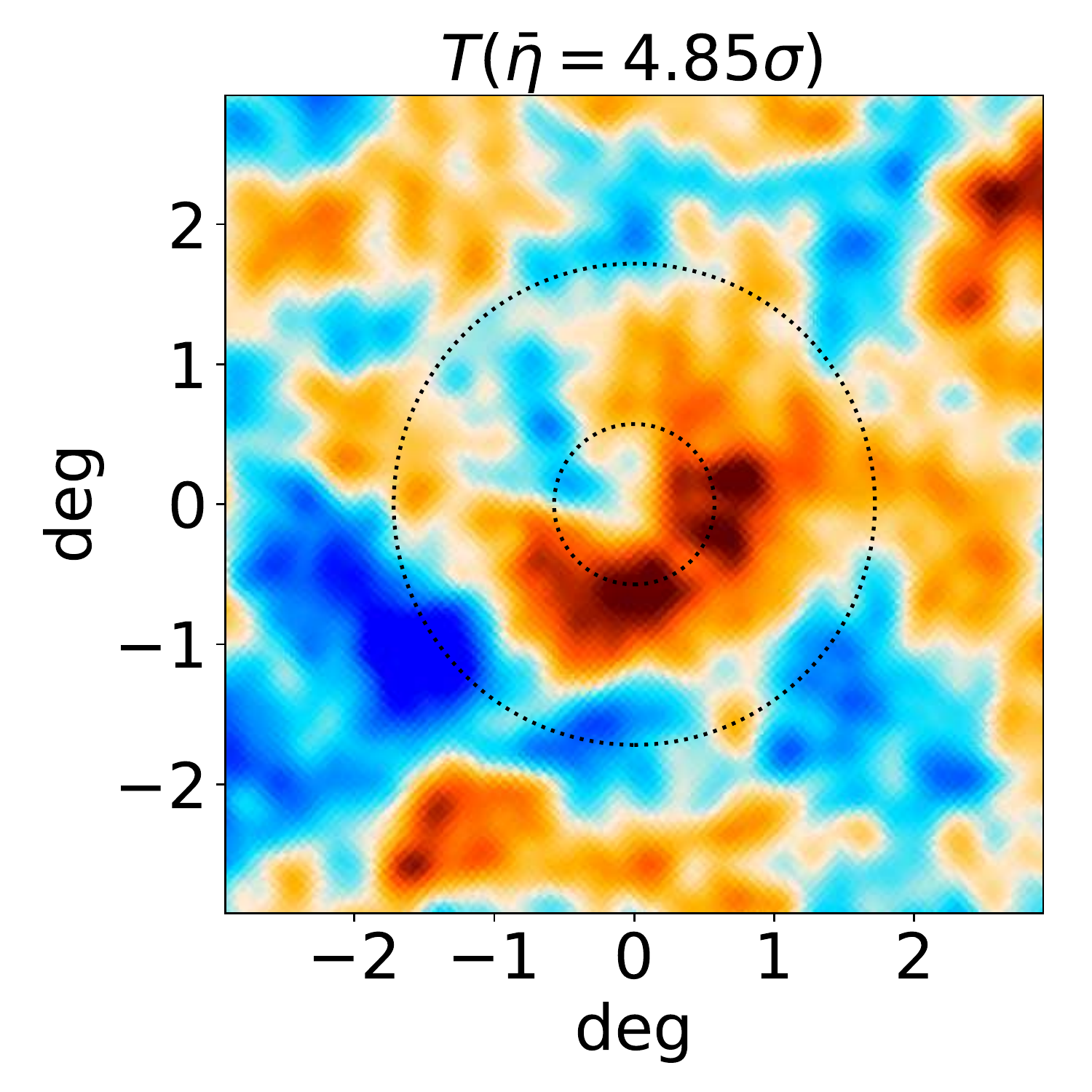}
\includegraphics[scale=0.3]{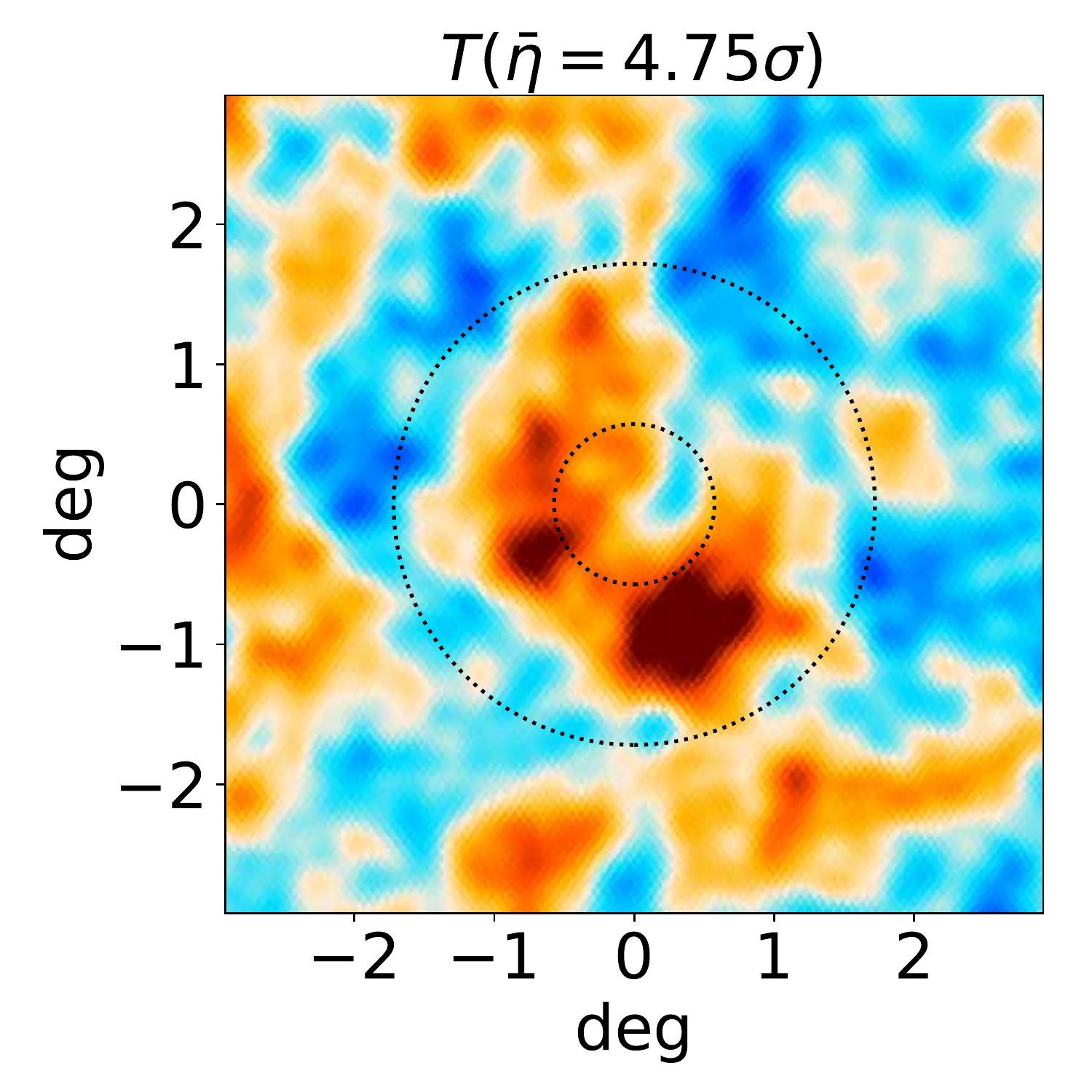}
\includegraphics[scale=0.3]{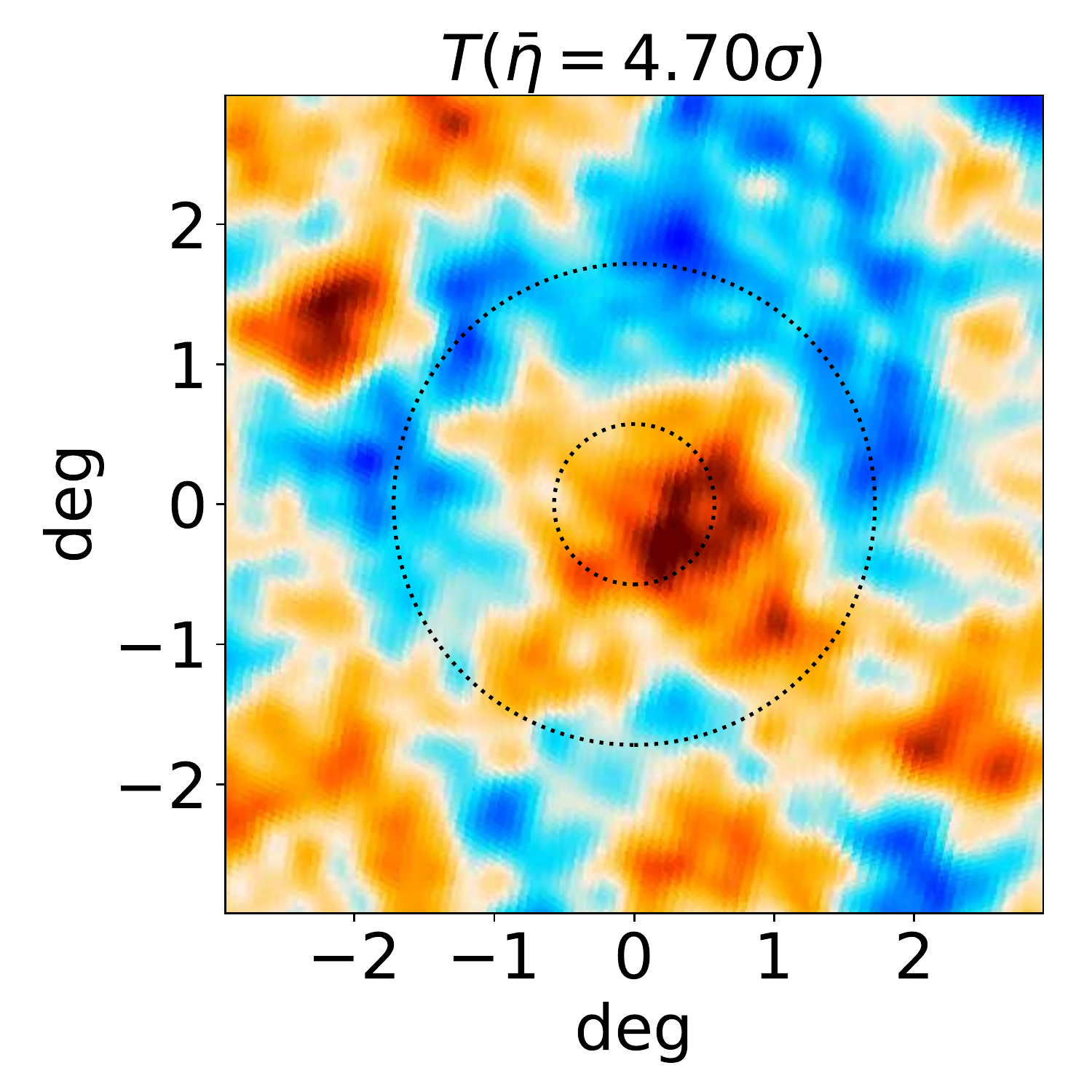} \\
\includegraphics[scale=0.3]{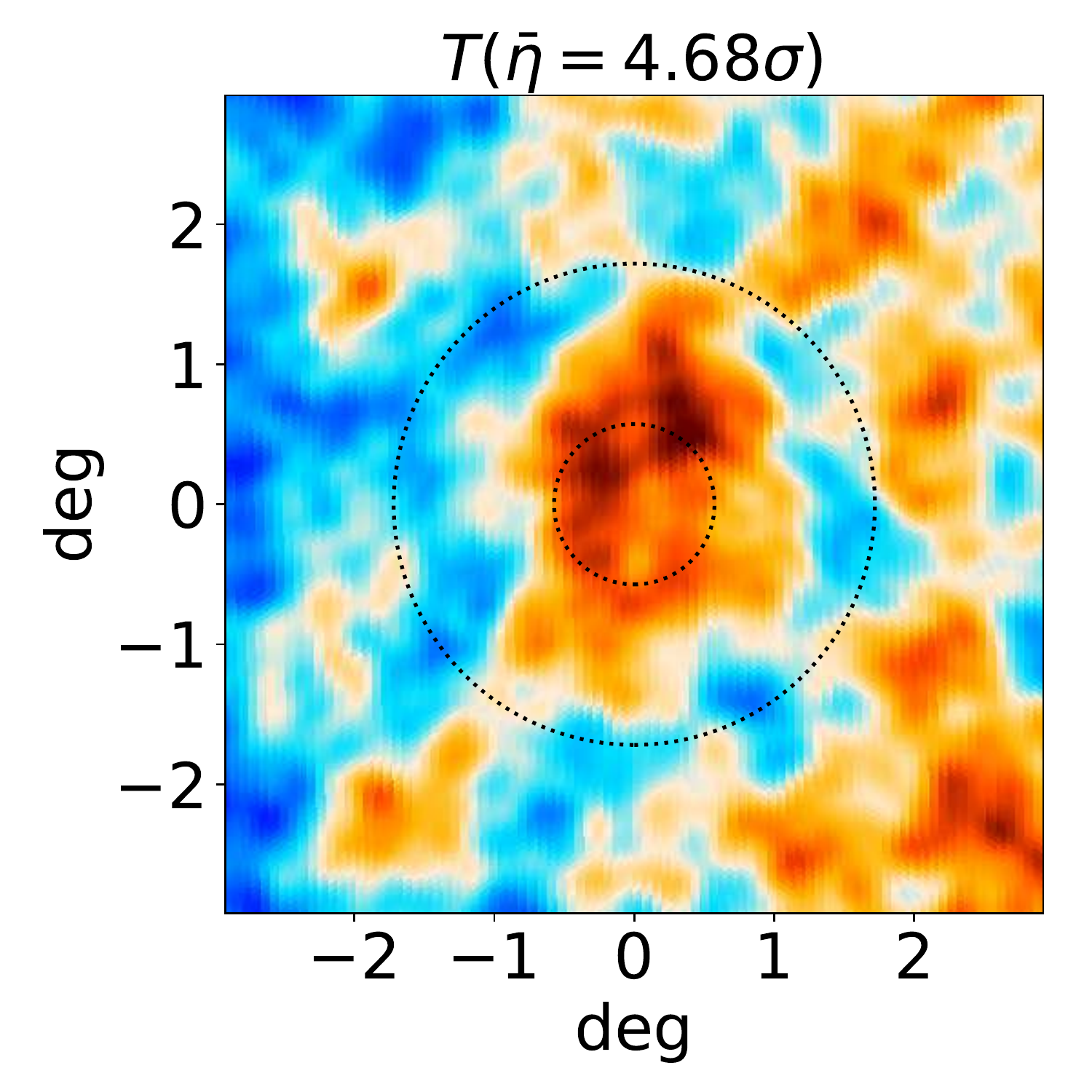}
\includegraphics[scale=0.3]{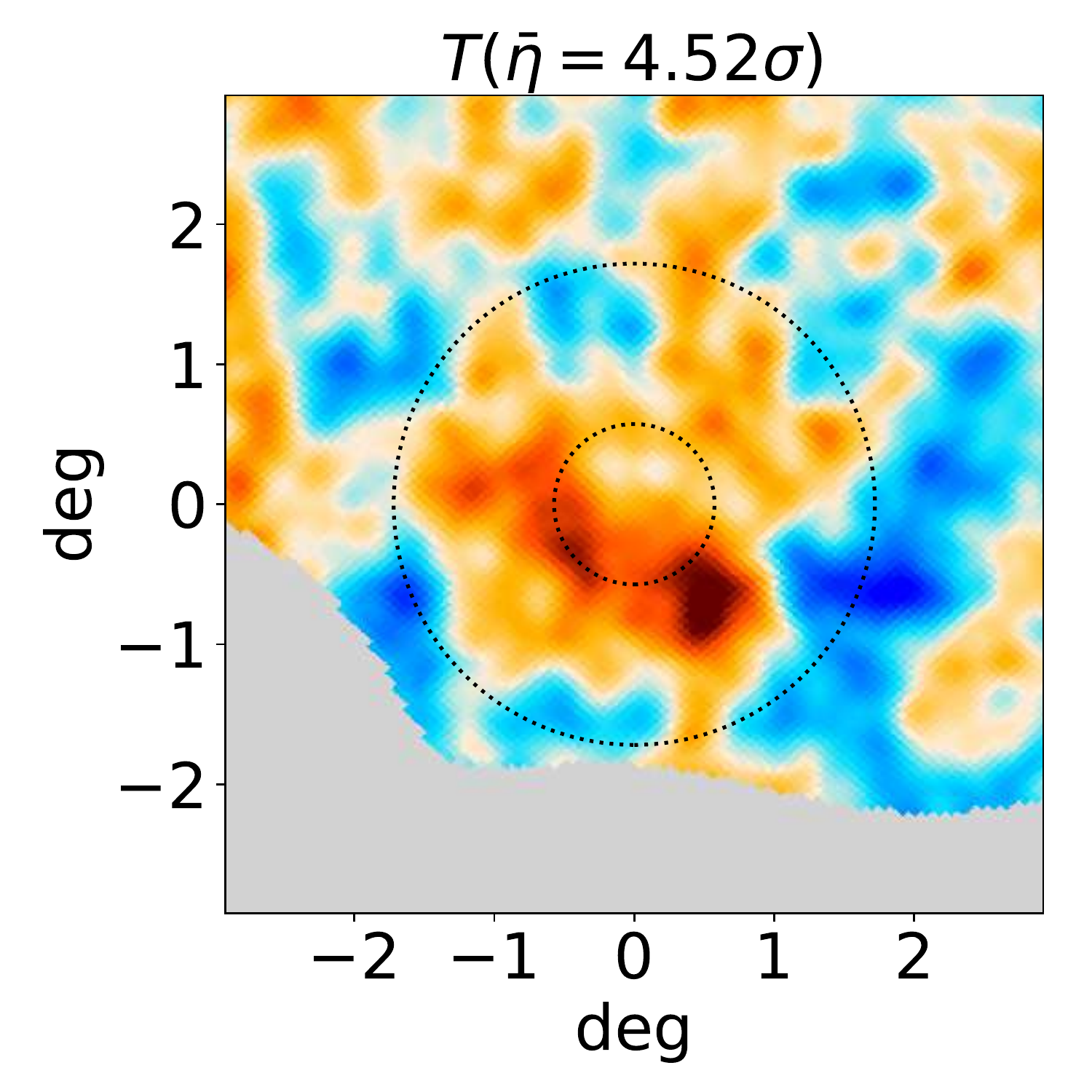}
\includegraphics[scale=0.3]{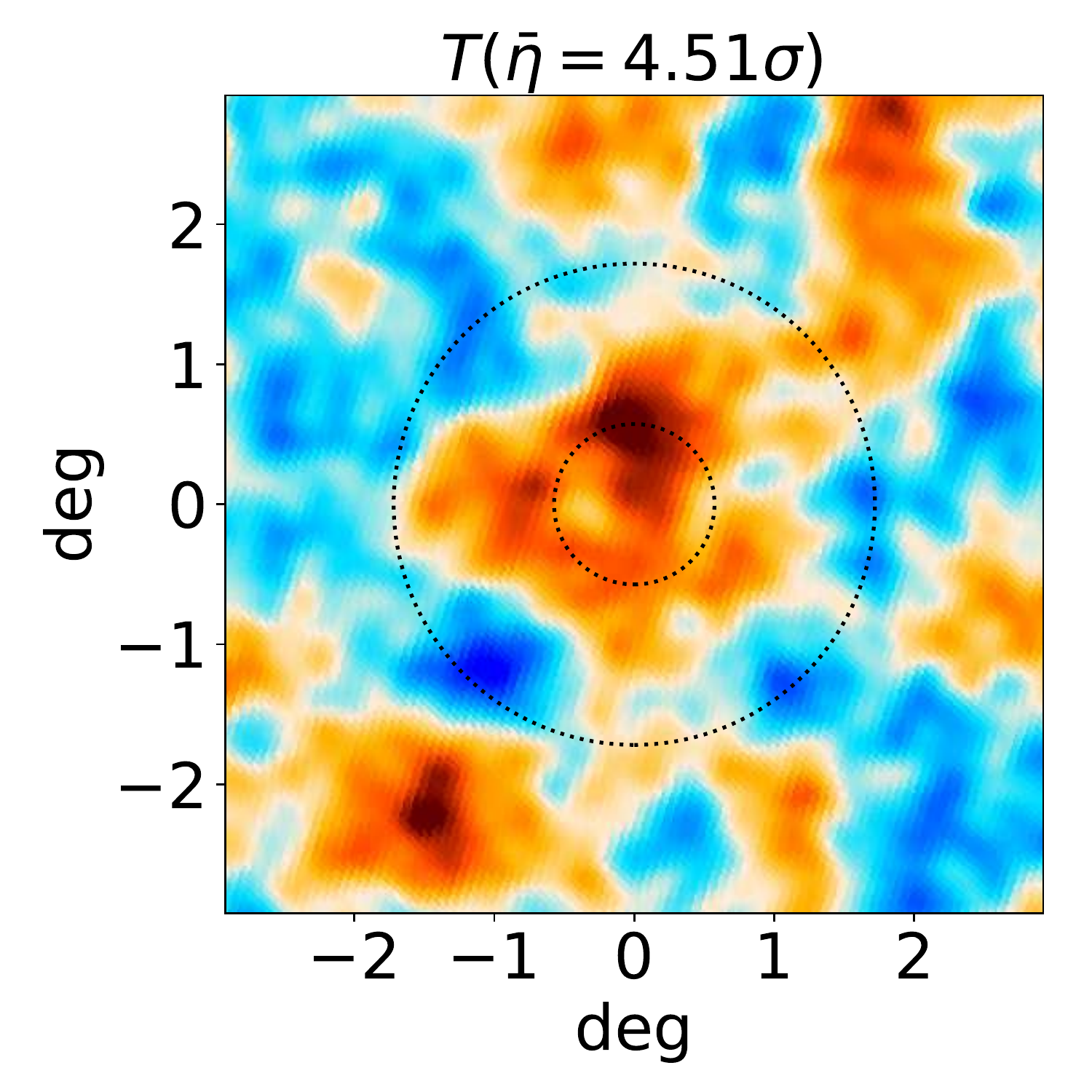} \\
\includegraphics[scale=0.3]{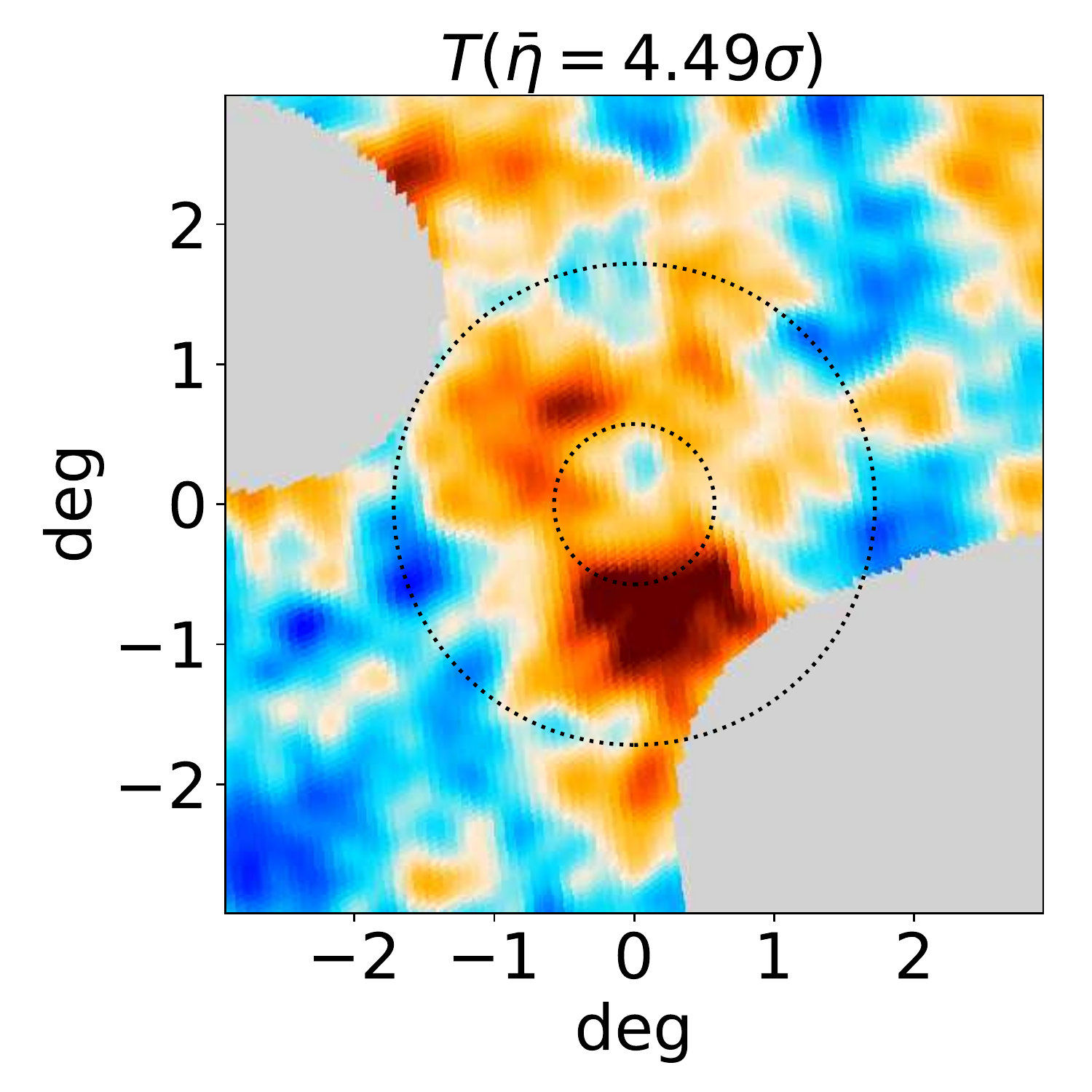}
\includegraphics[scale=0.3]{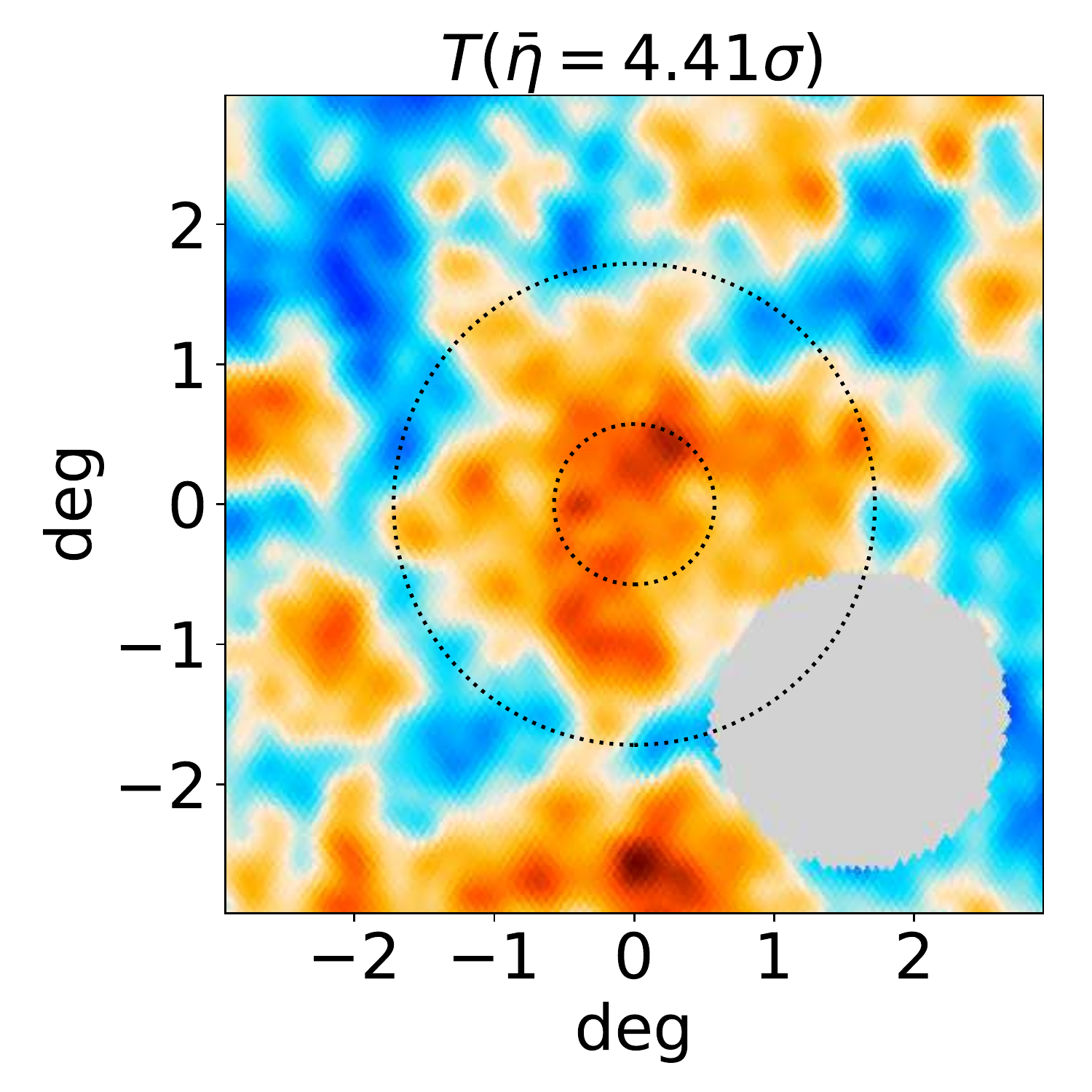}
\includegraphics[scale=0.3]{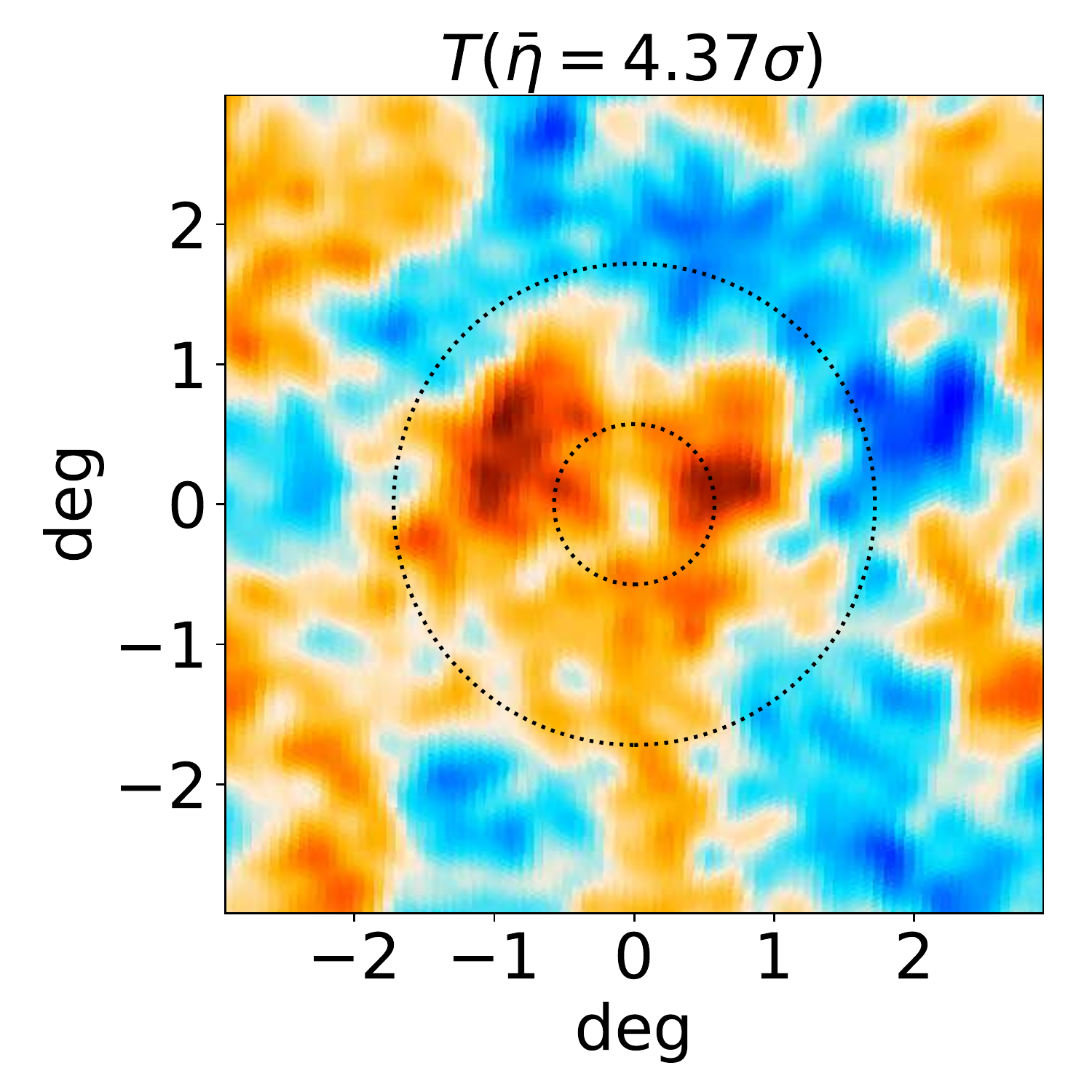} \\
\includegraphics[scale=0.3]{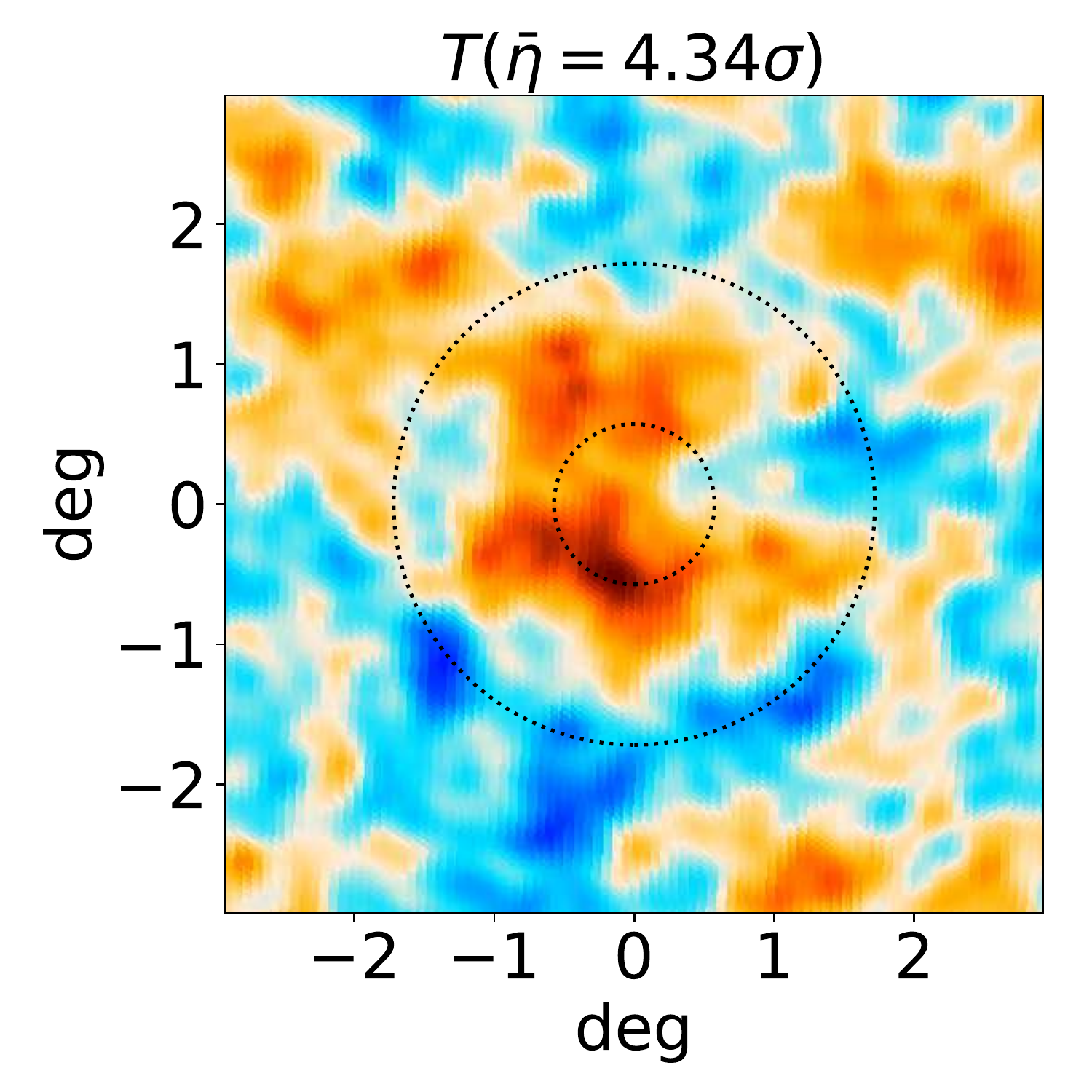}
\includegraphics[scale=0.3]{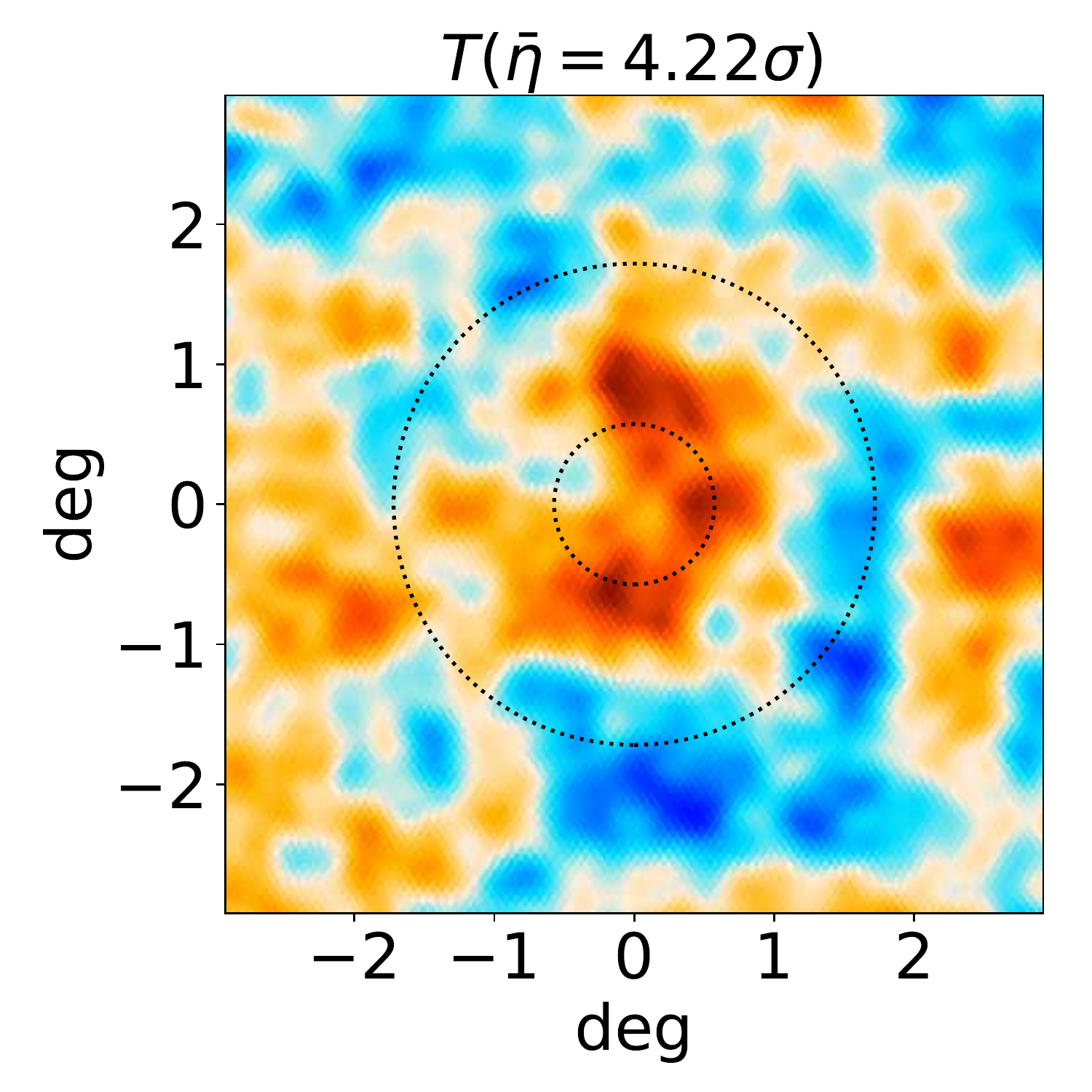}
\includegraphics[scale=0.3]{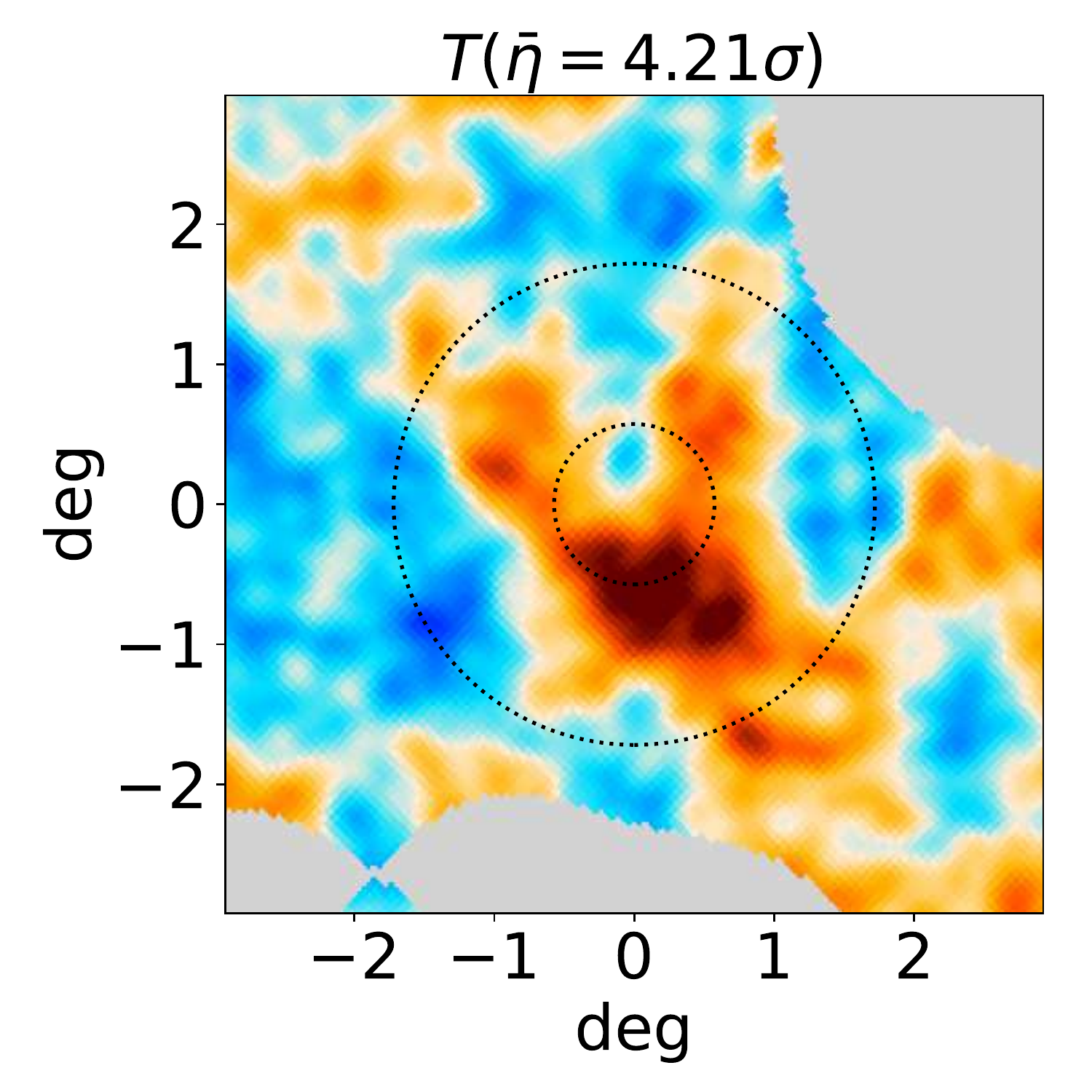}
}
\caption{CMB intensity patches centred on the sky directions in which the most prominent $15$ maxima are found in the normalized $\bar{\eta}$ map for a weight function with $r=0.01$~rad and $\epsilon=0.02$~rad. The mean value is subtracted from each patch for a better visualization, and the color scale covers from $-300$ to $300~\mu \mathrm{K}$. The grey regions depict data excluded by the extended \textit{Planck} confidence mask, while the dotted circumferences represent the edges of the region defined by $W$.}
\label{fig:patches}
\end{figure*}

In Figure~\ref{fig:patches}, we show the set of CMB intensity patches from the \sevem\ map centred on the sky directions depicted by red points in Figure~\ref{fig:directions}. In most of them, the azimuthal pattern is visually recognizable. But note that having a perfect ring is not necessary to obtain a maxima in $\bar{\eta}$, provided that there is an intense asymmetric contribution to the radial derivative throughout the region allowed by the corresponding $W$ function. Moreover, most of these patches present a positive excess in the central region. This is possibly due to the fact that, given the imposed constraint, the field has no room to vary within the disc defined by the inner radius of this particular $W$. Therefore, the limitations associated with this specific scale might accidentally entail that the low $p$-value is caused by a more general anomalous ``peakness'' at these particular scale and locations, having nothing to do with possible implications from CCC.

In addition, \citet{An2018hwk} also find a small $p$-value in the case of a weight function with $r=0.01$~rad and $\epsilon=0.03$~rad. No evidence of anomaly is found for maxima at this scale in the present analysis. Actually, a low $p$-value is obtained at this scale for the number of minima below $-3.5\sigma$, but the low probability does not persist at the most extreme thresholds. In any case, it would not correspond to the deviation from the model found by \citet{An2018hwk}, since in our analysis $\bar{\eta}$ minima represent directions in which the CMB temperature increases outwards. As shown in Section~\ref{sec:Elsewhere}, due to limitations to sample the tail of the extrema distribution, obtaining a probability of $0.002$ is not a real threat for the standard model.

\subsubsection*{Polarization analysis of the most deviated case}
\begin{figure}
\centering{
\includegraphics[scale=0.45]{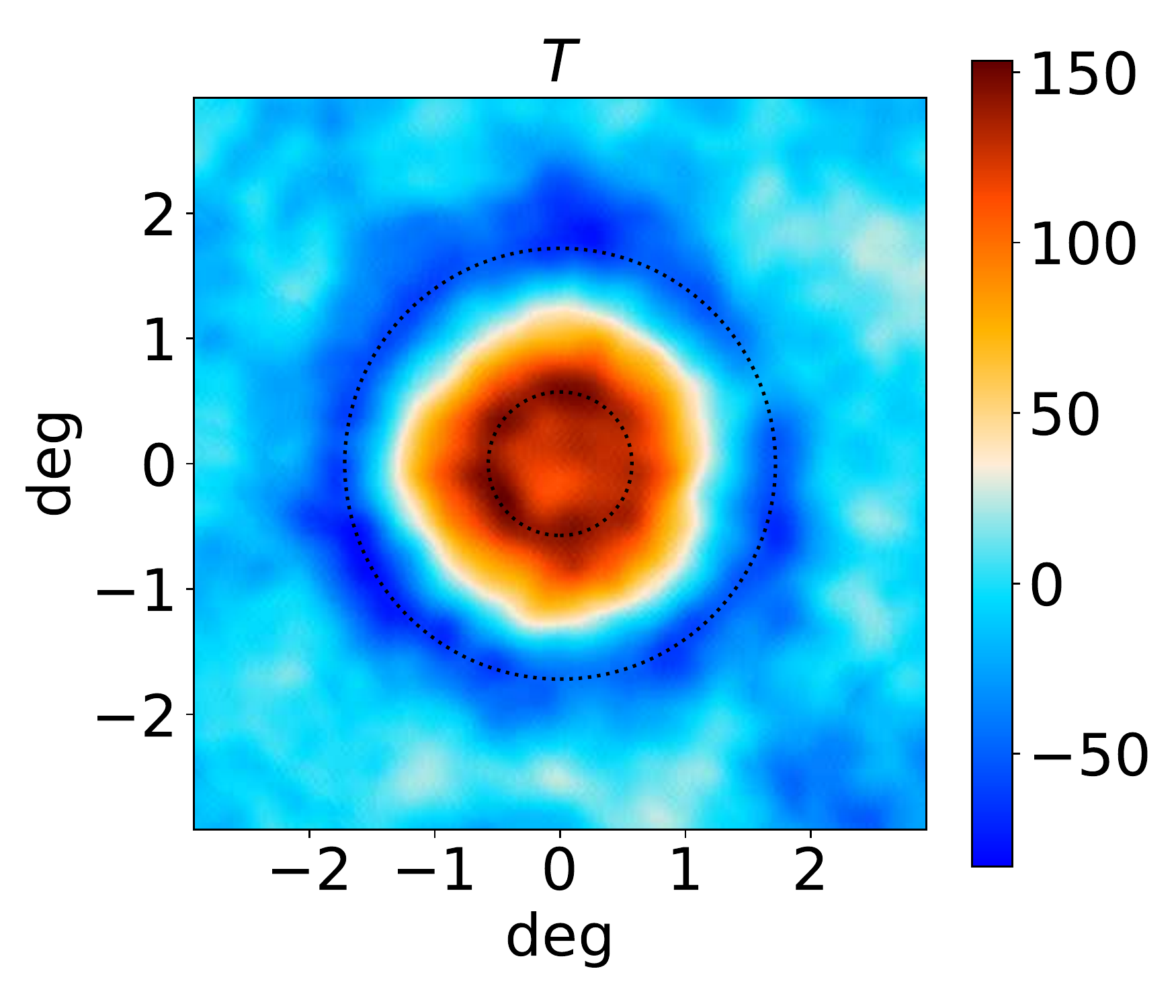} \\
\includegraphics[scale=0.45]{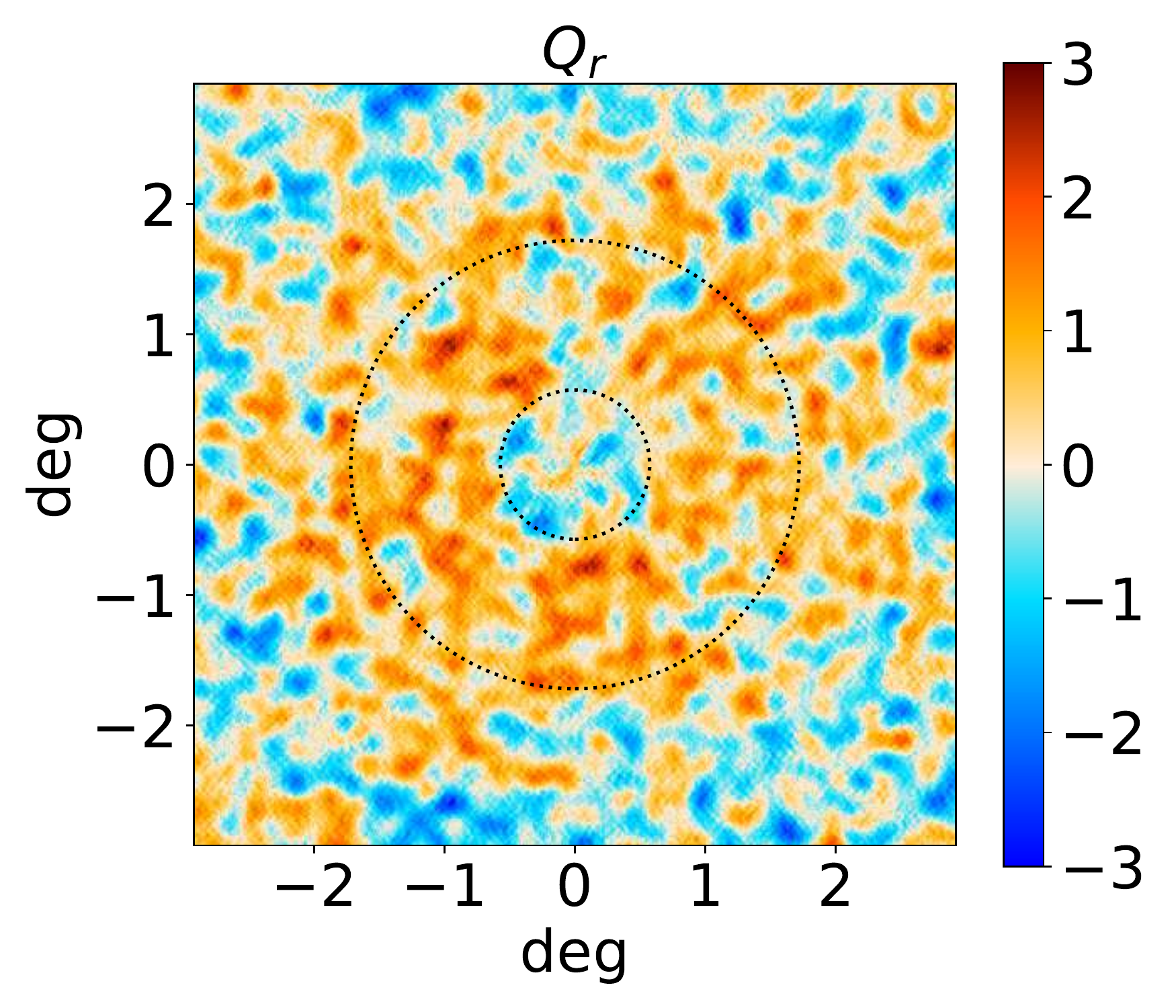} \\
\includegraphics[scale=0.45]{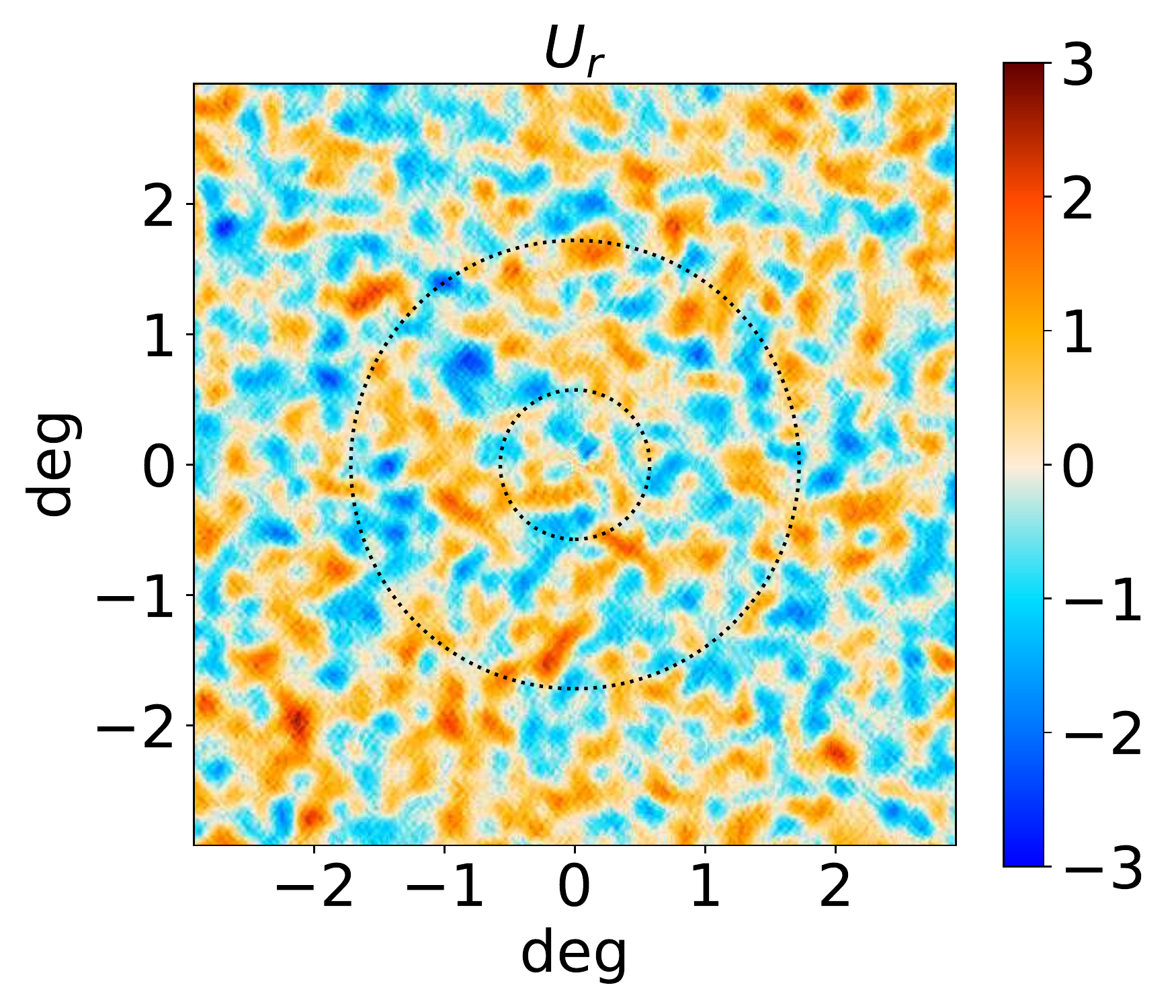} 
}
\caption{Stacked patches of CMB intensity (upper panel), $Q_r$ (middle panel) and $U_r$ (bottom panel) Stokes parameters centred at sky directions in which $\bar{\eta}$ maxima above $3.5\sigma$ are found with a weight function with $r=0.01$~rad and $\epsilon=0.02$~rad. The colour bars are expressed in $\mu \mathrm{K}$. The edges of the region defined by $W$ are depicted by the dotted circumferences.} 
\label{fig:patches_stack}
\end{figure}

\begin{figure*}
\centering{
\includegraphics[scale=0.6]{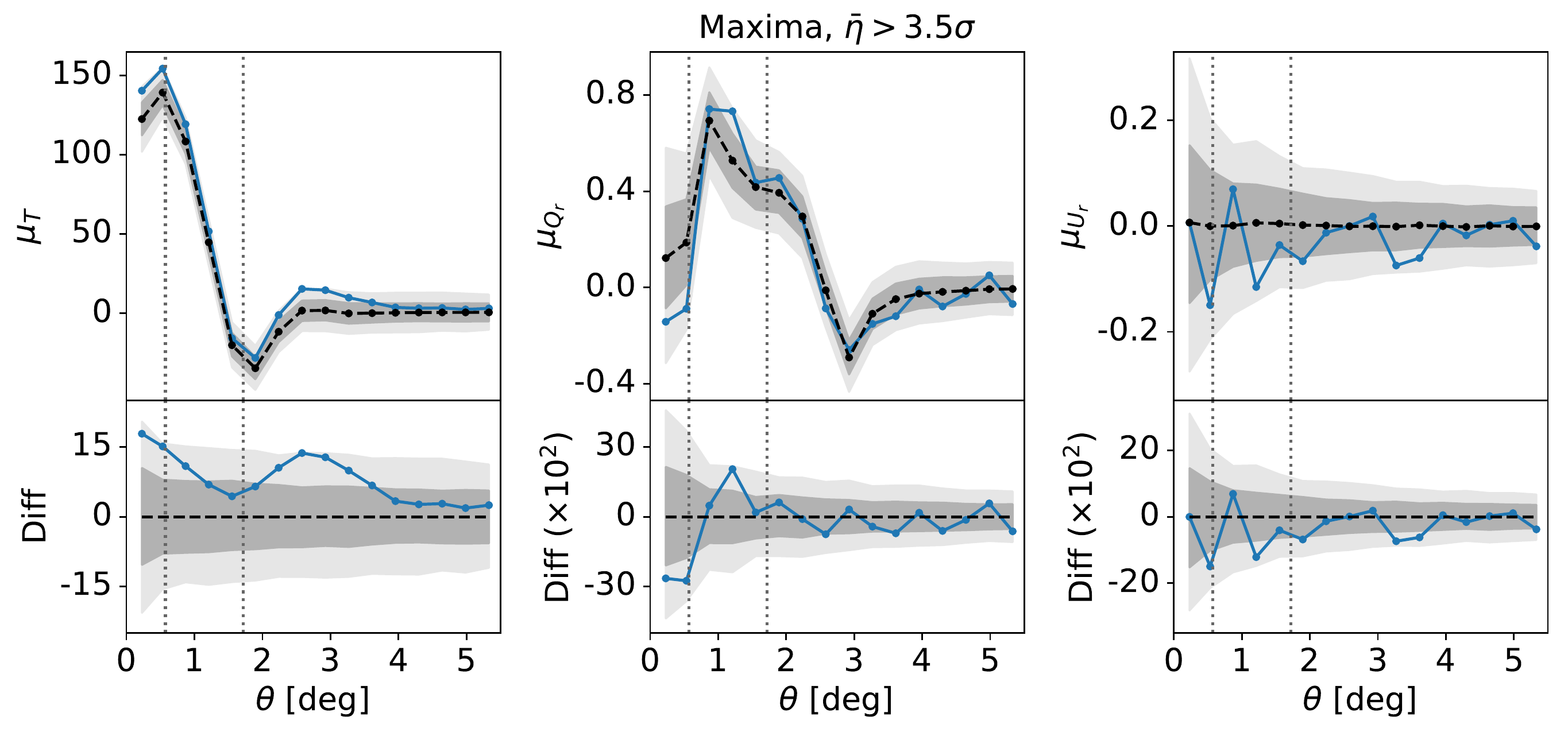} \\
\includegraphics[scale=0.6]{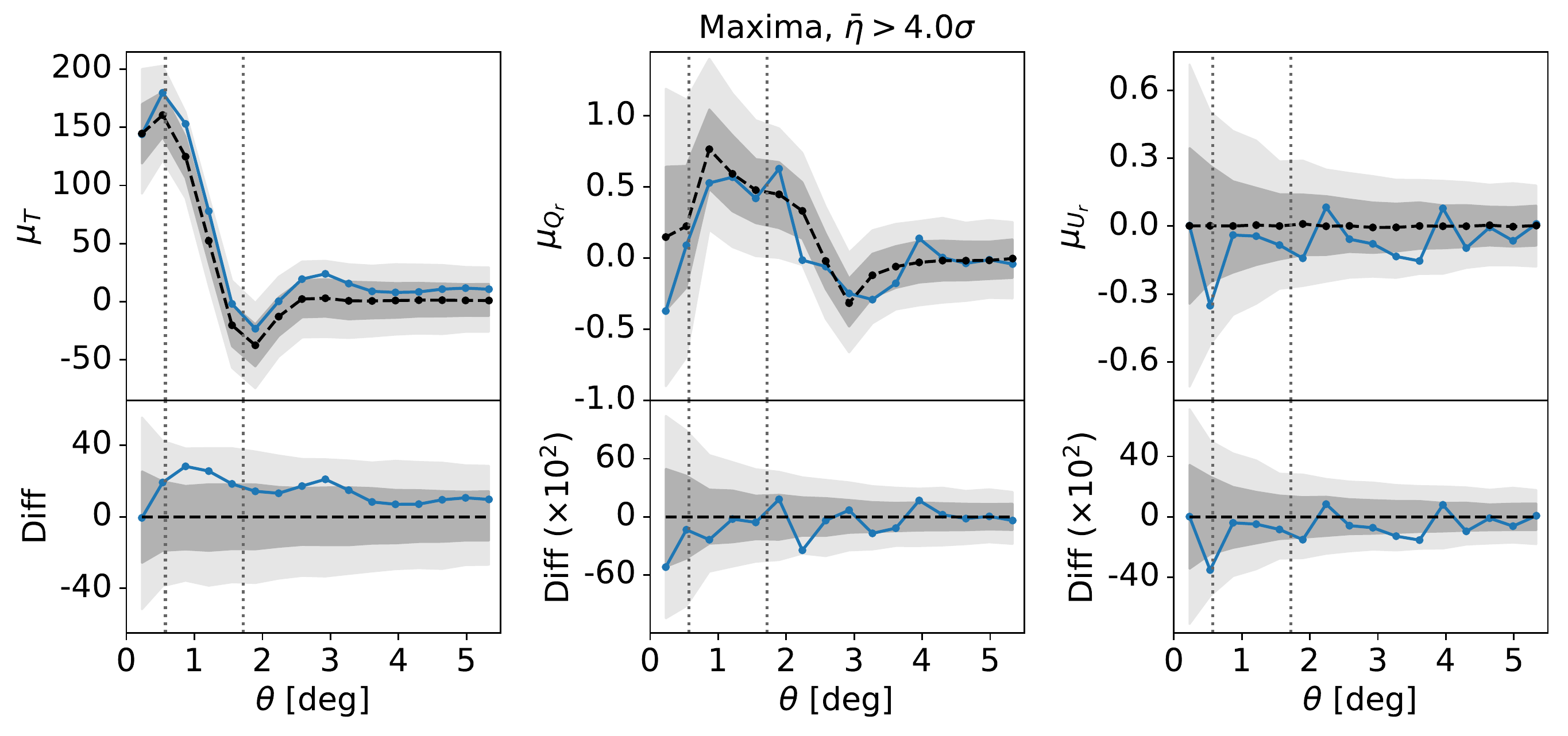}
}
\caption{Mean radial profiles in $\mu \mathrm{K}$ from the stacking of the CMB intensity and polarization at $\bar{\eta}$ maxima above $3.5\sigma$ (top panels) and $4.0\sigma$ (bottom panels) using a weight function with $r=0.01$~rad and $\epsilon=0.02$~rad. The blue line depicts the radial profile from the \sevem\ map, while the dashed black line is the mean value from FFP10 simulations. The shadowed regions show the $68$ and the $95$ per cent confidence levels estimated from simulations. The differences between data and simulations are explicity shown in the small bottom panels. The edges of the region defined by $W$ are depicted by the vertical dotted lines.}
\label{fig:profiles_stack}
\end{figure*}

The polarization pattern at the locations depicted in Figure~\ref{fig:directions} is also explored using the $Q_r$ and $U_r$ Stokes parameters. They provide a representation of the polarization vector in terms of a local frame in which the polarization axes are considered radially and tangentially with respect to a reference centre, such as those depicted in Figure~\ref{fig:spinor_components} \citep[see, for instance,][]{PlanckVII2018}. As the corresponding individual patches of $Q_r$ and $U_r$ are noisy, an alternative approach is applied. Let us consider, for the sake of the argument, that most of these objects are elements of a particular natural kind. Under this view, it would be convenient to perform a stacking analysis in order to increase the signal-to-noise ratio of their peculiar features. Under these circumstances, it may be possible to check if the polarization pattern is compatible with the expected one from the standard model. In this case, the radial profiles of $Q_r$ and $U_r$ depend on the angular cross-spectra between $\bar{\eta}$ and the corresponding polarization mode, namely $C^{\bar{\eta},E}_{\ell}$ and $C^{\bar{\eta},B}_{\ell}$, respectively \citep[see][]{MarcosCaballero2016}. Note that a contribution from $C^{\bar{\eta},E}_{\ell}$ is expected within the standard model. In fact, those cross-spectra are nothing else than filtered versions of the usual $C^{T,E}_{\ell}$ and $C^{T,B}_{\ell}$, respectively. Figure~\ref{fig:patches_stack} shows the stacked patches of CMB intensity, $Q_r$ and $U_r$ Stokes parameters centred at sky directions in which $\bar{\eta}$ maxima above $3.5\sigma$ are found with a weight function with $r=0.01$~rad and $\epsilon=0.02$~rad. The orientation of the patches is defined relative to their local meridian. In addition, it is shown a comparison between the angular radial profiles from the data and the model in Figure~\ref{fig:profiles_stack}. As expected, the CMB intensity patch presents an azimuthal pattern. In terms of the radial profile, the \textit{Planck} intensity presents a systematical deviation from the expected value. Nevertheless, the deviation is whithin the errorbars and we should take into account that the bins of the profile are expected to be highly correlated. We compute the following $\chi^2$ square from data and simulations:
\begin{equation}
\label{eq:chi2}
\chi^2 = \sum_{i,j=1}^n \left[\mu(\theta_i) - \bar{\mu}(\theta_i) \right] C^{-1}_{ij} \left[\mu(\theta_j) - \bar{\mu}(\theta_j)  \right], 
\end{equation}
where $i$ runs over $16$ rings at different angular distances from the center with an angular width of $20.6$ arcmin, and $\mu(\theta_i)$ is the mean angular profile of the selected extrema from the data at the center of each ring. The covariance matrix $C$ between different rings and the expected value of the angular profiles within the standard model $\bar{\mu}$ are computed from the FFP10 simulations. As the distribution of this estimator obtained from simulations fits well with a theoretical $\chi^2$ with $16$ degrees of freedom, we use it to compute a $p$-value as the probability of obtaining a $\chi^2$ value from a standard realization at least as great as the one computed from the data. In the case of the intensity profile, this yields a value of $0.037$ for maxima above $3.5\sigma$, which could be indicating a certain deviation. However, the $p$-value from maxima above $4.0\sigma$ is $0.841$. Regarding polarization, the $Q_r$ component from the data seems to be compatible with the predicted level within the standard model. As expected, the $U_r$ pattern is noisy for both data and simulations. The corresponding $p$-values for maxima above $3.5\sigma$ are $0.789$ and $0.553$ respectively, and for maxima above $4.0\sigma$ are $0.936$ and $0.866$. Therefore, we conclude that taking the group of extrema as a specific kind of events, the potentially anomalous effect that we could have measured is diluted. Note that for this $W$ function, $63$ independent $\bar{\eta}$ maxima are detected on the \sevem\ data above $3.5\sigma$, and $22$ above $4.0\sigma$. Only $9$ independent maxima are detected above $4.5\sigma$, but we are not able to compute properly the error bars of the profiles for this case as there are some realizations in which no maxima are obtained above this threshold.

\subsubsection*{Large-scale analysis}
Analogously, Table~\ref{tab:pvalues_extrema_gt_ls} shows the $p$-values computed from the number of extrema present in the $\bar{\eta}$ maps generated with different large-scale weight functions. As mentioned at the beginning of this section, due to the small number of extrema we keep in some cases, considering a subset of independent extrema is not useful for the large-scale analysis because we obtain an unreliable statistics. The most deviant case is then the counting of maxima above $3.5\sigma$ and $4\sigma$ for $r=0.26$~rad and $\epsilon = 0.04$~rad, which yields a probability of $0.001$. We also obtain a small $p$-value ($0.004$) for the case of maxima above $4\sigma$ for $r=0.06$~rad and $\epsilon = 0.04$~rad, but the fact that no deviation is observed above $4.5\sigma$ makes suspect that it could be a mere statistical fluke. In any case, as it is shown in Section~\ref{sec:Elsewhere}, a probability of $0.001$ is not small enough in the present analysis to be considered an anomaly.
 
Note that \citet{An2018} find the greatest deviation from a particular scale with a radius around $8^{\circ}$ when looking for differences between the mean value of the CMB intensity fluctuations within contiguous rings. This approach is a coarse-grain estimation of the radial derivative in the centred region of a major ring which contains the two contiguous ones. As there are some overlapping between those rings and the $W$ functions considered in the present analysis, it would be expected to observe a counterpart of that deviation from the model. However, it does not seem to be found in the present analysis. Nevertheless, motivated by the possibility that we might be overlooking relevant scales, we compute an extra case with $r = 0.08$~rad and $\epsilon = 0.12$~rad, that matches one of the scales (using their notation: $\gamma=0.14$, and $\epsilon = 0.06$) in Table 3 from \citet{An2018}. However, no anomalous $p$-values are obtained from this case either. 

One could argue that the fact that \citet{An2018} exclude all centers whose $W$ function overlaps a fraction more than $1$ per cent of the area with the confidence mask may increase the significance of their results. The size of the sample is drastically reduced when considering such an aggressive mask. For the smaller scales considered in the present analysis, this mask excludes typically an $80$-$90$ per cent of the sky. It rises above the $90$ per cent for radius greater than $0.30$~rad, making unfeasible the analysis of extrema for $r > 0.22$~rad even for the most permissive thresholds ($3.0\sigma$ and $3.5\sigma$). In any case, no small $p$-values are found for $r < 0.22$~rad excluding all centers whose $W$ function overlaps more than $1$ per cent with the \textit{Planck} confidence mask, not even for $r=0.06$~rad and $\epsilon = 0.04$~rad.

\begin{table*}
\begin{center}
\resizebox{\textwidth}{!} {
\begin{tabular}{|cc|cccc|cccc|}
\hline
\multicolumn{2}{|c|}{Weight functions} & \multicolumn{4}{c|}{$P$-values for maxima} & \multicolumn{4}{c|}{$P$-values for minima} \\
$r$ (rad) & $\epsilon$ (rad)  & $>3.0\sigma$ & $>3.5\sigma$ & $>4.0\sigma$ & $>4.5\sigma$ & $<-3.0\sigma$ & $<-3.5\sigma$ & $<-4.0\sigma$ & $<-4.5\sigma$\\
\hline
\hline
\multirow{3}{*}{0.06} & 0.04 & $0.217$ & $0.044$ & $0.004$ & $0.147$ & $0.806$ & $0.506$ & $0.437$ & $0.340$ \\
					  & 0.08 & $0.815$ & $0.606$ & $0.140$ & $0.042$ & $0.931$ & $0.803$ & $0.503$ & $0.413$ \\
					  & 0.12 & $0.791$ & $0.849$ & $0.668$ & $0.286$ & $0.671$ & $0.196$ & $0.136$ & $0.060$ \\
\hline
\multirow{3}{*}{0.10} & 0.04 & $0.497$ & $0.258$ & $0.174$ & $0.159$ & $0.937$ & $0.907$ & $0.756$ & $0.146$ \\
					  & 0.08 & $0.512$ & $0.873$ & $0.853$ & $0.407$ & $0.411$ & $0.240$ & $0.118$ & $0.021$ \\
					  & 0.12 & $0.686$ & $0.381$ & $0.547$ & $0.272$ & $0.312$ & $0.104$ & $0.089$ & $0.164$ \\
\hline					  
\multirow{3}{*}{0.14} & 0.04 & $0.472$ & $0.324$ & $0.307$ & $0.091$ & $0.298$ & $0.453$ & $0.739$ & $0.144$ \\
					  & 0.08 & $0.404$ & $0.218$ & $0.240$ & $0.478$ & $0.439$ & $0.429$ & $0.402$ & $0.196$ \\
					  & 0.12 & $0.631$ & $0.481$ & $0.636$ & $0.286$ & $0.699$ & $0.550$ & $0.882$ & $0.268$ \\
\hline
\multirow{3}{*}{0.18} & 0.04 & $0.252$ & $0.054$ & $0.696$ & $0.326$ & $0.580$ & $0.617$ & $0.348$ & $0.168$ \\
					  & 0.08 & $0.770$ & $0.628$ & $0.659$ & $0.421$ & $0.978$ & $0.818$ & $0.396$ & $0.404$ \\
					  & 0.12 & $0.801$ & $0.782$ & $0.893$ & $0.297$ & $0.861$ & $0.962$ & $0.863$ & $0.303$ \\
\hline
\multirow{3}{*}{0.22} & 0.04 & $0.632$ & $0.634$ & $0.296$ & $0.334$ & $0.798$ & $0.928$ & $0.994$ & $0.328$ \\
					  & 0.08 & $0.759$ & $0.453$ & $0.254$ & $0.414$ & $0.757$ & $0.718$ & $0.744$ & $0.422$ \\
					  & 0.12 & $0.344$ & $0.122$ & $0.086$ & $0.101$ & $0.591$ & $0.662$ & $0.754$ & $0.268$ \\
\hline
\multirow{3}{*}{0.26} & 0.04 & $0.038$ & $0.001$ & $0.001$ & $0.084$ & $0.780$ & $0.374$ & $0.244$ & $0.079$ \\
					  & 0.08 & $0.157$ & $0.052$ & $0.114$ & $0.120$ & $0.780$ & $0.859$ & $0.820$ & $0.400$ \\
					  & 0.12 & $0.273$ & $0.087$ & $0.061$ & $0.033$ & $0.766$ & $0.764$ & $0.543$ & $0.062$ \\
\hline
\multirow{3}{*}{0.30} & 0.04 & $0.827$ & $0.473$ & $0.614$ & $0.166$ & $0.832$ & $0.550$ & $0.594$ & $0.334$ \\
					  & 0.08 & $0.559$ & $0.329$ & $0.549$ & $0.201$ & $0.904$ & $0.692$ & $0.117$ & $0.426$ \\	
					  & 0.12 & $0.661$ & $0.464$ & $0.456$ & $0.283$ & $0.756$ & $0.753$ & $0.721$ & $0.256$ \\
\hline
\multirow{3}{*}{0.34} & 0.04 & $0.552$ & $0.408$ & $0.702$ & $0.618$ & $0.679$ & $0.670$ & $0.774$ & $0.143$ \\
					  & 0.08 & $0.871$ & $0.708$ & $0.638$ & $0.232$ & $0.603$ & $0.753$ & $0.718$ & $0.404$ \\
					  & 0.12 & $0.766$ & $0.551$ & $0.420$ & $0.149$ & $0.542$ & $0.626$ & $0.847$ & $0.268$ \\
\hline
\end{tabular}}
\end{center}
\caption{Probability of finding a number of $\bar{\eta}$ extrema in simulations which is strictly greater than that obtained from the \sevem\ data for different thresholds and large-scale weight functions with inner radius $r$ and thickness $\epsilon$.}
\label{tab:pvalues_extrema_gt_ls}
\end{table*}

\subsection{Cumulative distribution functions}
\label{subsec:CDF_sec}
In this Section, we consider an alternative statistic, which is also sensitive to the tails of the CDF from the normalized $\bar{\eta}$ map. In particular, it was also followed by \citet{An2018} and \citet{An2018hwk} to evaluate their results. Specifically, the tails of the CDF from data and simulations are compared by using the estimator described in \citet{Meissner2012}:
\begin{eqnarray}
\label{eq:AR}
A_R & = & -\dfrac{a}{N}\sum_{i=1}^n {d_i \ln{\left[ 1 - \left( F(x_i)  \right)^a\right]}}, \nonumber \\
A_L & = & -\dfrac{a}{N}\sum_{i=1}^n {d_i \ln{\left[ 1 - \left( 1 - F(x_i)  \right)^a\right]}},
\end{eqnarray}
where $N$ is the total number of unmasked pixels in the $\bar{\eta}$ map. It is assumed that the CDF is sampled using $n$ bins, in such a way that $x_i$ denotes the centre of the $i^\mathrm{th}$ bin and $d_i$ represents the number of  points contained in the corresponding bin. A theoretical CDF, $F$, is computed as the average of the CDFs obtained from the first $900$ FFP10 simulations. The estimator is parameterized by a positive real number $a$, which changes the weight of the tails. Given a large enough value for $a$, the outcome of the present analysis does not significantly depend on its value. To the extend that we are interested in comparing with \citet{An2018hwk}, the results of this section are shown for $a = 10000$, which is the value they use. Typically, $99\%$ of the value of $A_{R}$ ($A_{L}$) is determined by bins with $\bar{\eta} \gtrsim 3.3\sigma$ ($\bar{\eta} \lesssim -3.3\sigma$), while $80\%$ comes from bins with $\bar{\eta} \gtrsim 3.5\sigma$ ($\bar{\eta} \lesssim -3.5\sigma$). For instance, taking $n = 10000$, almost all the contribution comes from $\sim 800$ bins for each tail (while $80\%$ of the contribution comes from $\sim 400$ bins). Within reasonable values, the total number of bins is not important either. Finally, we considered a $p$-value computed as $P(A_{R/L} \geq A_{R/L}^{\mathrm{data}})$, namely the probability of finding in standard-model realizations a value of $A_{R/L}$ at least as great as the value obtained from the data.

For the small scales, the $p$-values associated with the $A_L$ and $A_R$ estimators for $a=10000$ and different $W$ functions are shown in Table~\ref{tab:pvalues}. In addition, the histograms of the $A_{L/R}$ values from simulations are shown in Figure~\ref{fig:histograms}, where the vertical lines represent the value from the data. As in the case of the number of extrema, the major deviation from the model is found using a weight function with $r=0.01$~rad and $\epsilon=0.02$~rad. This $p$-value is consistently low for $a > 1000$. Ultimately, a great value of $A_R$ is consistent with a greater number of maxima with large values of $\bar{\eta}$.  

\begin{figure*}
\includegraphics[scale=0.45]{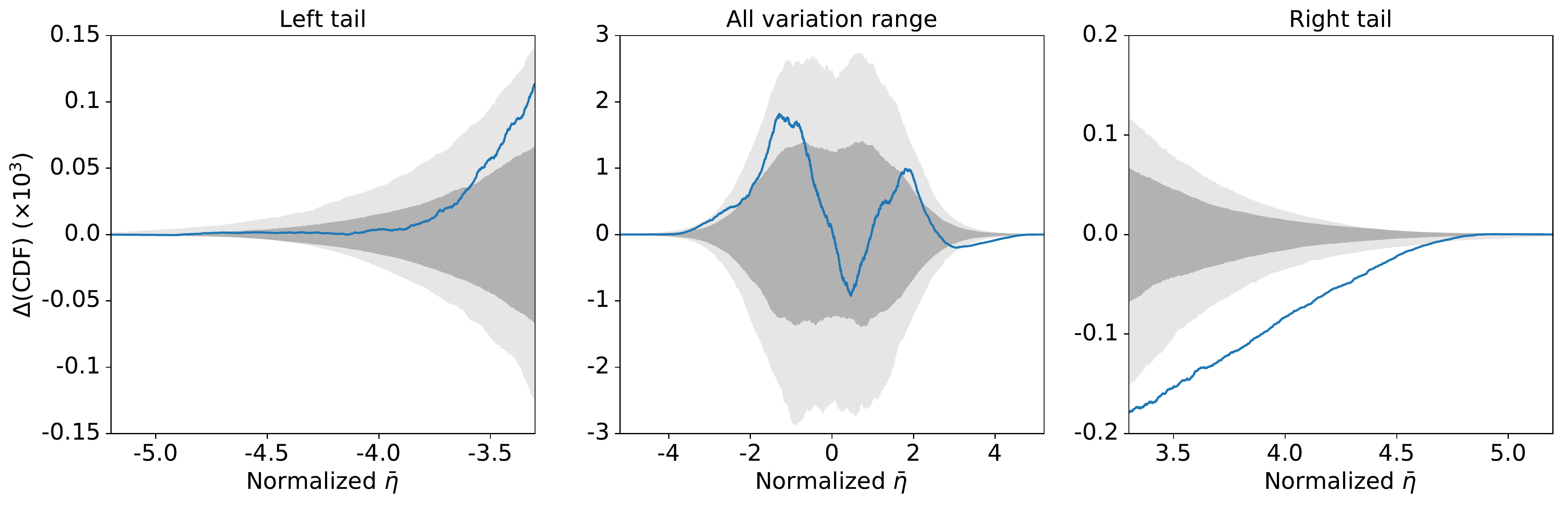}
\caption{Difference between the CDF from the \sevem\ data and the theoretical CDF computed as the mean from the FFP10 simulations for a $W$ function with $r=0.01$~rad and $\epsilon=0.02$~rad. The whole range is shown in the middle panel, while a zoom of the left and right tails is shown in the left and right panels respectively. The shaded regions correspond to the $68\%$ and $95\%$ confidence levels estimated from the FFP10 simulations.}
\label{fig:CDF}
\end{figure*}

Note that the estimators described by Eq.~\ref{eq:AR} are only valid within the domain in which the theoretical CDF is not saturated. This means that this approach excludes some data points in those cases where the tails of the CDF from the data are more extended than the tails of the theoretical CDF. This should not be particularly dramatic as, in these cases, the valid data points in the right (left) tail would be systematically lower (higher) than the theoretical ones. For the particular case with $r=0.01$~rad and $\epsilon=0.02$~rad, we observe that the range of variation of the CDF from the data is within the domain of the theoretical CDF when centres are considered at $\mathrm{N_{side}} = 1024$. However, this is not always the case, since the left tail from the data is more extended than the theoretical one when centres are selected from a grid at $\mathrm{N_{side}} = 64$. The former case is shown in Figure~\ref{fig:CDF}, in which the difference between the CDF from data and simulations is plotted (central panel). A zoom on the left (left panel) and right (right panel) tails is also shown. Note that the variation scale for the left tail is lower than the variation for the right one. The right tail of the CDF from data is systematically lower than the simulated one up to $\bar{\eta} \sim 4.8 \sigma$, although it should be kept in mind that these points are highly correlated. Some points are then concentrated around this value so that the CDF from data saturates before the theoretical one.

\begin{table}
\begin{center}
\resizebox{0.45\textwidth}{!} {
\begin{tabular}{|cc|cc|}
\hline
\multicolumn{2}{|c|}{Weight functions} & \multicolumn{2}{c|}{$P$-values} \\
$r$ (rad) & $\epsilon$ (rad)  & $A_L$ & $A_R$ \\
\hline
\hline
\multirow{4}{*}{0.00} & 0.01 & $0.832$ & $0.734$ \\
					  & 0.02 & $0.997$ & $0.336$ \\ 
					  & 0.03 & $0.286$ & $0.008$ \\
					  & 0.04 & $0.312$ & $0.598$ \\	
\hline
\multirow{4}{*}{0.01} & 0.01 & $0.949$ & $0.030$ \\
					  & 0.02 & $0.249$ & $<0.001$ \\ 
					  & 0.03 & $0.328$ & $0.321$ \\
					  & 0.04 & $0.382$ & $0.853$ \\	
\hline
\multirow{4}{*}{0.02} & 0.01 & $0.117$ & $0.034$ \\
					  & 0.02 & $0.570$ & $0.084$ \\ 
					  & 0.03 & $0.412$ & $0.830$ \\
					  & 0.04 & $0.440$ & $0.969$ \\	
\hline
\multirow{4}{*}{0.03} & 0.01 & $0.351$ & $0.282$ \\
					  & 0.02 & $0.777$ & $0.360$ \\ 
					  & 0.03 & $0.159$ & $0.636$ \\
					  & 0.04 & $ 0.201$ & $0.483$ \\	
\hline
\multirow{4}{*}{0.04} & 0.01 & $0.049$ & $0.281$ \\
					  & 0.02 & $0.037$ & $0.168$ \\ 
					  & 0.03 & $0.317$ & $0.674$ \\
					  & 0.04 & $0.017$ & $0.454$ \\	
\hline
\end{tabular}
}
\end{center}
\caption{Given the \textit{Planck} model, the probability of finding a value of $A_{R/L}$ with $a = 10000$ at least as great as the value obtained from the \sevem\ data for different small-scale weight functions with inner radius $r$ and thickness $\epsilon$.}
\label{tab:pvalues}
\end{table}

Finally, the $p$-values computed from the CDF for the corresponding large-scale $W$ functions are shown in Table~\ref{tab:pvalues_ls}. The most deviated value is $0.993$ for $A_L$ with $r=0.22$~rad and $\epsilon=0.04$~rad. Keeping only those centers whose $W$ functions overlap less than $1$ per cent with the confidence mask, the lowest $p$-values are $0.007$ for $A_L$ with $r=0.10$~rad and $\epsilon=0.08$~rad, and $0.009$ for $A_L$ with $r=0.06$~rad and $\epsilon=0.12$~rad, which are not small enough to be considered anomalous (as shown in the next section).

\begin{table}
\begin{center}
\resizebox{0.45\textwidth}{!} {
\begin{tabular}{|cc|cc|}
\hline
\multicolumn{2}{|c|}{Weight functions} & \multicolumn{2}{c|}{$P$-values} \\
$r$ (rad) & $\epsilon$ (rad)  & $A_L$ & $A_R$ \\
\hline
\hline
\multirow{3}{*}{0.06} & 0.04 & $0.357$ & $0.063$ \\
					  & 0.08 & $0.566$ & $0.049$ \\
					  & 0.12 & $0.076$ & $0.831$ \\
\hline					  
\multirow{3}{*}{0.10} & 0.04 & $0.720$ & $0.286$  \\
					  & 0.08 & $0.031$ & $0.947$ \\ 
					  & 0.12 & $0.159$ & $0.561$ \\
\hline
\multirow{3}{*}{0.14} & 0.04 & $0.432$ & $0.384$ \\
					  & 0.08 & $0.279$ & $0.104$ \\ 
					  & 0.12 & $0.798$ & $0.634$ \\
\hline
\multirow{3}{*}{0.18} & 0.04 & $0.381$ & $0.640$ \\
					  & 0.08 & $0.519$ & $0.796$ \\ 
					  & 0.12 & $0.988$ & $0.814$ \\
\hline
\multirow{3}{*}{0.22} & 0.04 & $0.993$ & $0.410$ \\
					  & 0.08 & $0.843$ & $0.417$ \\ 
					  & 0.12 & $0.797$ & $0.091$ \\
\hline
\multirow{3}{*}{0.26} & 0.04 & $0.190$ & $0.029$ \\
					  & 0.08 & $0.896$ & $0.179$ \\ 
					  & 0.12 & $0.317$ & $0.081$ \\
\hline
\multirow{3}{*}{0.30} & 0.04 & $0.535$ & $0.496$ \\
					  & 0.08 & $0.282$ & $0.430$ \\ 
					  & 0.12 & $0.667$ & $0.470$ \\
\hline
\multirow{3}{*}{0.34} & 0.04 & $0.441$ & $0.650$ \\
					  & 0.08 & $0.792$ & $0.572$  \\ 
					  & 0.12 & $0.807$ & $0.410$ \\
\hline
\end{tabular}
}
\end{center}
\caption{Given the \textit{Planck} model, the probability of finding a value of $A_{R/L}$ with $a = 10000$ at least as great as the value obtained from the data for different large-scale weight functions with inner radius $r$ and thickness $\epsilon$.}
\label{tab:pvalues_ls}
\end{table}

%

\section[The significance problem]{The significance problem}
\label{sec:Elsewhere}
\begin{figure*}
\includegraphics[scale=0.28]{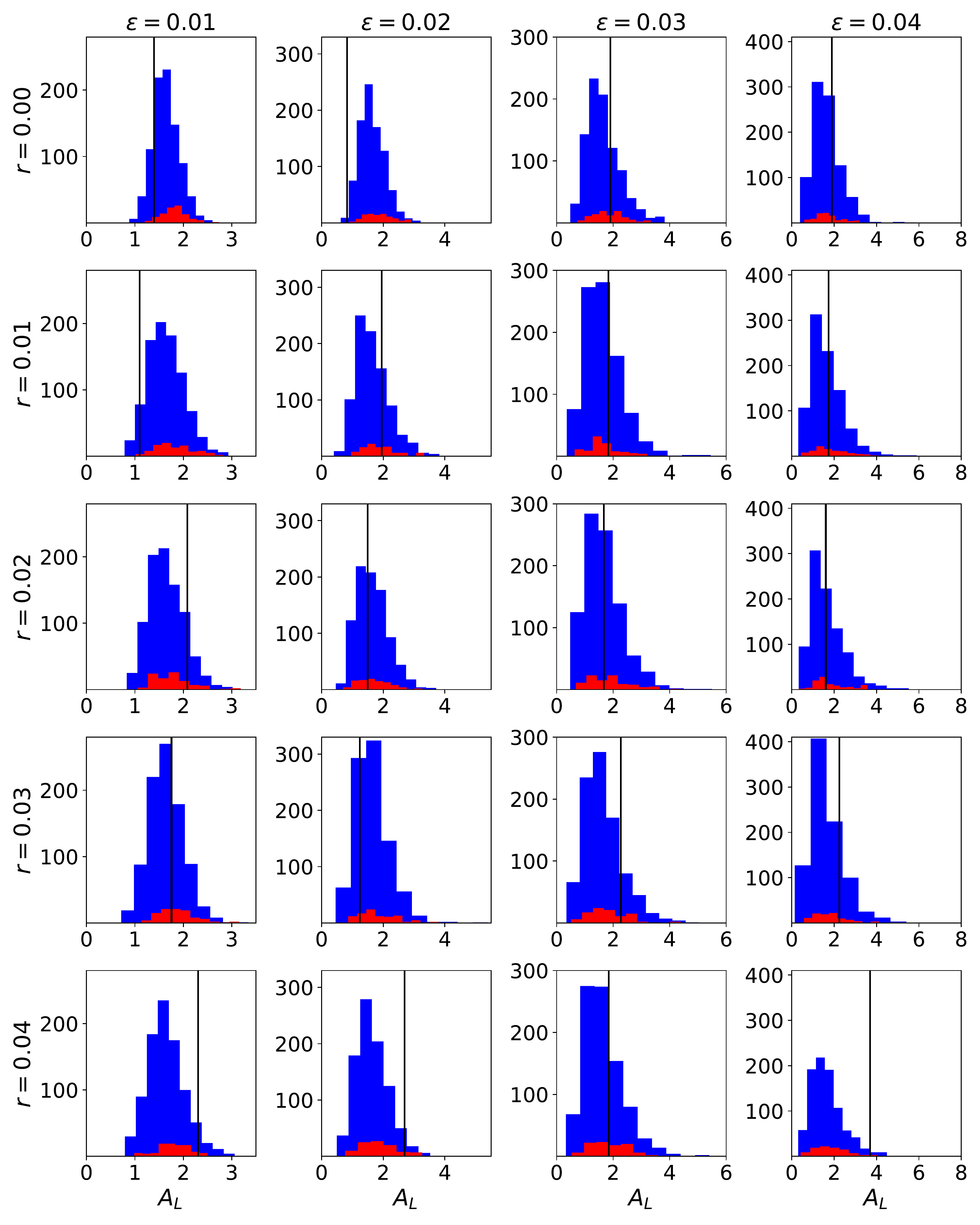}
\includegraphics[scale=0.28]{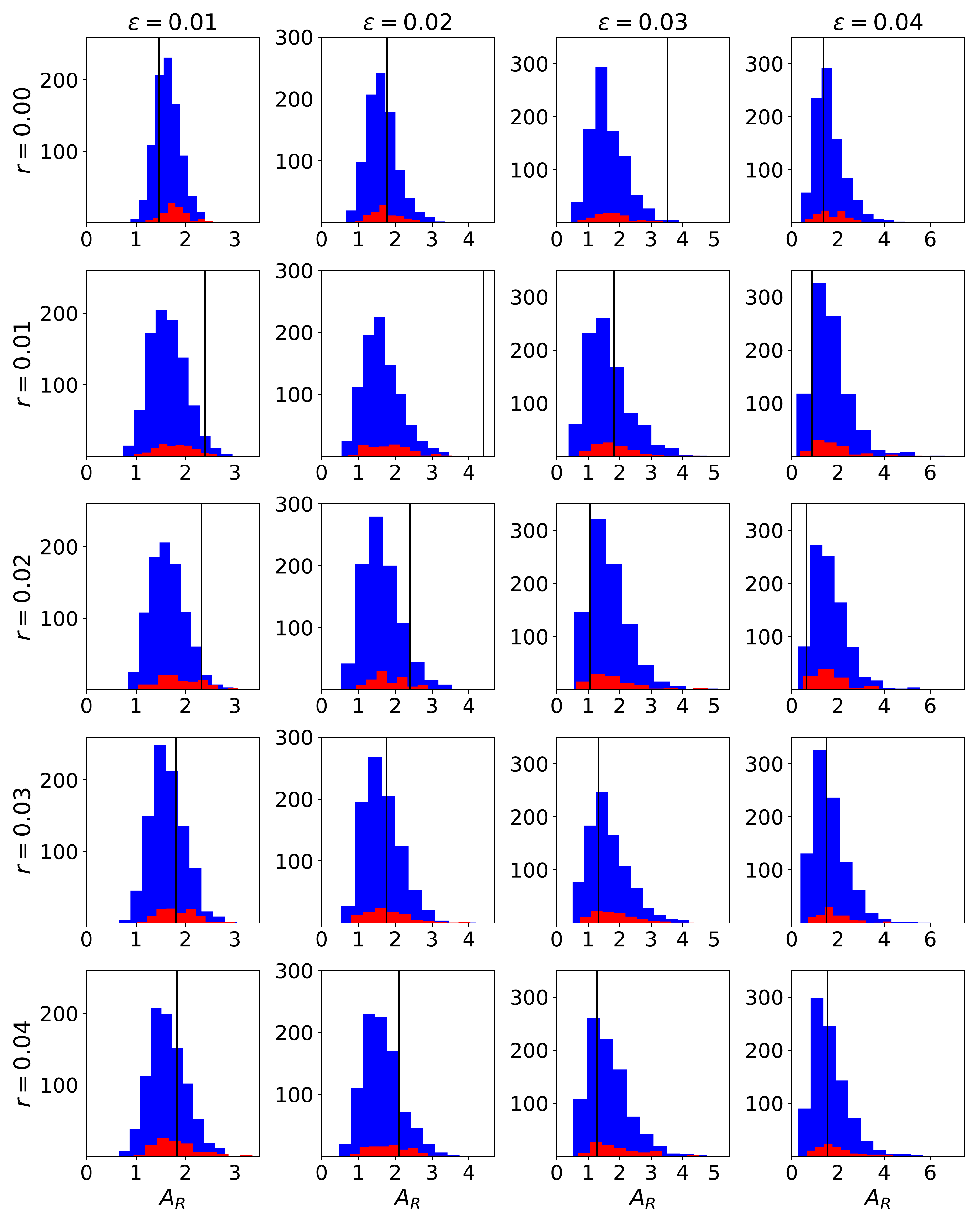}
\caption{Values of $A_L$ (left panels) and $A_R$ (right panels) from data and simulations for different small-scale weight functions parameterized by the inner radius $r$ and the thickness $\epsilon$. The blue histograms depict the values from the first $900$ FFP10 simulations, while the red histograms show the subset of the last $99$ FFP10 simulations. The values from the \sevem\ data are represented by the black vertical lines.}
\label{fig:histograms}
\end{figure*}

No FFP10 realization as deviated as the data has been found when convolving the normalized $\bar{\eta}$ map with a weight function with $r=0.01$~rad and $\epsilon=0.02$~rad. Despite this evidence, it is hard to determine what is the statistical significance of the observed deviation from the standard prediction. In the first place, we are only able to assign an upper limit to the probability ($<1/1000$) of finding such value for the radial derivative. As an example, the histograms tracing the probability distribution of $A_{L/R}$ values in the standard model for different weight functions are shown in Figure~\ref{fig:histograms} (in blue). In addition, we use the last $99$ FFP10 simulations as an independent set of standard realizations. Their $A_{L/R}$ values are shown as the red histograms. The black vertical lines depict the values obtained from the \sevem\ data map. As in a few cases the red histograms contain $A_{L/R}$ values greater than the whole set obtained from the first $900$ simulations, it is obvious that this number of simulations is not enough to trace properly the right tail of the probability distributions. An analogous problem is also present when analysing the number of extrema. 

In the second place, we should be aware that the whole configuration space in this test involves different scales (explicitly, $20$ for the small-scale, and $24$ for the large-scale analyses) for the two tails of the CDFs (or alternatively, for the number of two types of extrema above four different thresholds). So, given the mentioned limitations of the significance test, it is important to estimate how likely are such deviations in this larger configuration space. To check the possibility of a look-elsewhere effect, we use the $99$ extra simulations which were excluded in the previous section as independent data sets. As the $p$-value is saturated when it is obtained a value of $A_{L/R}$ greater than those computed from the first $900$ realizations, we consider another approach to evaluate how unexpected are the results. In particular, an asymmetric distance is defined in terms of the width which enclose a $68\%$ of the area under each distribution around the median value. The deviation of the data from the median is then considered in units of the corresponding distance (towards the left or the right, depending on the position of the data with respect to the median). For the small-scale analysis, we find that $3$ of $99$ simulations present a deviation from the median of the $A_{L}$ or $A_{R}$ values which is greater, for at least one of the considered scales, than the one obtained for the real data in the case with $r=0.01$~rad and $\epsilon=0.02$~rad. 

The analysis of the number of extrema is a bit different, because the highest thresholds are dominated by a Poissonian regime. Assuming that the counting of extrema is described by a Poissonian distribution, we are able to compute its parameter $\lambda$ as the mean of the number computed from the $900$ simulations. Therefore, we define the characteristical distance for each distribution in terms of $\sqrt{\lambda}$. In this case, the deviation of the data is then computed from the $\lambda$ value and expressed in $\sqrt{\lambda}$ units. In the small-scale analysis, when the $99$ simulations convolved by any of the considered $W$ functions are explored, we find that $2$ of them (both for a threshold of $4.5\sigma$) present a greater deviation than the one observed in the data for $r=0.01$~rad and $\epsilon=0.02$~rad when considering the whole set of extrema. On the contrary, no simulation is found with a greater deviation when considering the subset of independent extrema. However, the selection procedure causes a depopulation in the histogram tails, where the frequencies are low beforehand. This affects the calculation of $\sqrt{\lambda}$, making the deviation appear artificially larger. Moreover, for the large scales, we find many realizations with greater deviations than the one associated with the lowest $p$-values in Table~\ref{tab:pvalues_extrema_gt_ls}.

Therefore, in view of all these results, the small $p$-values quoted in Section~\ref{sec:Analysis} should be interpreted as statistical flukes.

%

\section[Conclusions]{Conclusions}
\label{sec:Conclusions}
We have reviewed some of the most recent results in the literature regarding the presence of ring-type structures in the CMB fluctuations such as those predicted by the CCC. In particular, a methodology based on the radial derivative is applied to the \textit{Planck} data. As it can be reduced to a CMB map filtering, this approach is much faster than other methods computed in real space.

Our analysis is splitted in two distinct regimes exploring both small and large scales. In each one of them, several $W$ functions are considered, testing different values for the inner radius and the thickness of the ring. Additionally, for each weight function, two different estimators are applied to the \textit{Planck} data and compared with their expected values from realistic FFP10 simulations. On the one hand, we use the counting of extrema in the averaged radial derivative $\bar{\eta}$ map to check if there is an unusual presence of sky directions with anomalously great radial derivatives. On the other hand, the comparison is made in terms of the tails of the CDF from the normalized $\bar{\eta}$ map.

At small scales, the most pronounced deviation from the model is obtained from the same angular scale than the one  pointed by \citet{An2018hwk}. However, in our assessment, and consistently with \citet{Jow2020}, the confidence level when considering this deviation to be anomalous is significantly smaller than the $99.98\%$ claimed by these authors. Although our results are limited by the finite number of simulations, specially in relation to the look-elsewhere effect analysis, they show no strong evidence to claim a significant deviation from the standard model. Regarding the large-scale analysis, we conclude that the statistical significance of the deviations from the model are not large enough to be considered anomalous.



\section*{acknowledgments}
The authors would like to thank Spanish Agencia Estatal de Investigaci\'on (AEI, MICIU) for the financial support provided under the projects with references ESP2017-83921-C2-1-R and AYA2017-90675-REDC, co-funded with EU FEDER funds, and also acknowledge the funding from Unidad de Excelencia Mar\'ia de Maeztu (MDM-2017-0765). AM-C acknowledges the postdoctoral contract from the University of the Basque Country UPV/EHU ``Especializaci\'on de personal investigador doctor'' program, and the financial support from the Spanish Ministry MINECO, MCIU/AEI/FEDER grant (PGC2018-094626-B-C21), the Basque Government grant (IT979-16). This research used resources of the National Energy Research Scientific Computing Center (NERSC), a U.S. Department of Energy Office of Science User Facility operated under Contract No. DE-AC02-05CH11231.

\section*{Data availability}
The data and simulations underlying this article are available in Planck Legacy Archive, at http://pla.esac.esa.int/pla/.

\bibliographystyle{mn2e}
\bibliography{citas_ringtype}

\end{document}